\documentclass[aps,prd,10pt,twocolumn,superscriptaddress,showpacs,footinbib]{revtex4-1}
\pdfoutput=1
\usepackage{hyperref}
\usepackage{slashed}
\usepackage{bm}
\usepackage{graphicx}
\usepackage{enumerate}
\usepackage{amsfonts,amsmath,amssymb,bm,bbm}
\usepackage{color}
\usepackage{epsfig, subfigure}
\usepackage{setspace}
\usepackage{booktabs, tabularx} 
\usepackage{multirow}

\hypersetup{
    pdfnewwindow=true,      % links in new window
    colorlinks=true,       % false: boxed links; true: colored links
    linkcolor=blue,          % color of internal links
    citecolor=blue,        % color of links to bibliography
    filecolor=blue,      % color of file links
    urlcolor=blue        % color of external links
}

\newcommand{\be}{\begin{equation}}  
\newcommand{\ee}{\end{equation}}
\newcommand{\bea}{\begin{eqnarray}}  
\newcommand{\eea}{\end{eqnarray}}

\begin{document}

\title{Direct detection of dark matter in universal bound states}

\author{Ranjan Laha}
\affiliation{Center for Cosmology and AstroParticle Physics (CCAPP), Ohio State University, Columbus, OH 43210}
\affiliation{Department of Physics, Ohio State University, Columbus, OH 43210\\
{\tt laha.1@osu.edu,braaten@mps.ohio-state.edu}\smallskip}

\author{Eric Braaten}
\affiliation{Department of Physics, Ohio State University, Columbus, OH 43210\\
{\tt laha.1@osu.edu,braaten@mps.ohio-state.edu}\smallskip}

\date{\today}

%%%%%%%%%%%%%%%%%%%%%%%%%%%%%%%%%%%%%%%%%%%%%%%%%%%
\begin{abstract}
We study the signatures for internal structure of dark matter in direct detection experiments in the context of asymmetric self-interacting dark matter.  The self-interaction cross section of two dark matter particles at low energies is assumed to come close to saturating the S-wave unitarity bound, which requires the presence of a resonance near their scattering threshold.  The universality of S-wave near-threshold resonances then implies that the low-energy scattering properties of a two-body bound state of dark matter particles are completely determined by its binding energy, irrespective of the underlying microphysics.  The form factor for elastic scattering of the bound state from a nucleus and the possibility of breakup of the bound state produce new signatures in the nuclear recoil energy spectrum.  If these features are observed in experiments, it will give a smoking-gun signature for the internal structure of dark matter.
\end{abstract}
%%%%%%%%%%%%%%%%%%%%%%%%%%%%%%%%%%%%%%%%%%%%%%%%%%%
%\pacs{95.35.+d, 14.60.Lm, 14.60.St, 98.80.-k} 

\maketitle

%%%%%%%%%%%%%%%%%%%%%%%%%%%%%%%%%%%%%%%%%%%%%%%%%%%
\section{Introduction}
\label{sec:introduction}
%%%%%%%%%%%%%%%%%%%%%%%%%%%%%%%%%%%%%%%%%%%%%%%%%%%

The presence of dark matter in the Universe has been inferred gravitationally for the last $\sim$ 80 years.  However, in spite of decades of search, we do not know the particle content of the dark sector.  Among the many prospective candidates for dark matter, a massive neutral particle is favored as the  dark matter candidate for many compelling theoretical reasons.  Search for the particle properties of dark matter proceeds via direct detection, indirect detection, production in colliders, and the search for effects in galaxy formation.

In spite of the enormous success of the $\Lambda$CDM model in explaining the observations of the large-scale structures in our Universe, several small-scale anomalies\,\cite{Weinberg:2013aya} (missing satellites\,\cite{Kravtsov:2009gi}, core vs.\,cusp\,\cite{Walker:2012td} and too big to fail\,\cite{BoylanKolchin:2011de}) have called for a modification of the collisionless dark matter paradigm\,\cite{Spergel:1999mh,Rocha:2012jg,Peter:2012jh}.  Although the possibility of baryonic feedback being a solution to these problems is not yet completely excluded\,\cite{Pontzen:2011ty,Governato:2012fa},  several particle physics models have been built to incorporate strong self-interactions among the dark matter particles\,\cite{Feng:2009mn,Feng:2009hw,Buckley:2009in,Loeb:2010gj,Tulin:2012wi,Tulin:2013teo,Fan:2013yva,Fan:2013tia,Dasgupta:2013era,Khlopov:2013ava}.

Asymmetric dark matter is mainly motivated by the observation that the present day dark matter density and the baryon density differ only by a factor of $\sim$ 5.  In the early Universe, the Sakharov conditions created an asymmetric mixture of baryons and antibaryons.  The present baryon density is the remnant after all the antibaryons have annihilated away.  It is possible that the Sakharov conditions also created an asymmetry between the particles and antiparticles of dark matter in the early universe.  This requires the dark matter particle to be distinct from its antiparticle.   Generally, the dark matter particle in asymmetric dark matter models are light, but exceptions exist.  The present dark matter could be a remnant after all the antiparticles have annihilated away. The generation mechanisms of the dark matter density and the baryon density may be related in asymmetric dark matter models\,\cite{Petraki:2013wwa,Zurek:2013wia}. 

Much of the present baryonic matter in the universe consists of particles with internal structure.  Protons and neutrons are composed of quarks.  Nuclei are bound states of protons and neutrons.  An atom is a bound state of a nucleus and electrons.  Dark matter is most often assumed to consist of individual elementary particles.  However it is possible that some or all of the present dark matter consists of particles with internal structure.  Internal structure of dark matter has been discussed in the context of enhanced annihilation cross sections required to explain the positron excess\,\cite{MarchRussell:2008tu}.  The search for bound states of weakly interacting dark matter particles in colliders has also been proposed in Ref.\,\cite{Shepherd:2009sa}.

A new way of looking at some dark matter properties was recently pointed out in Ref.\,\cite{Braaten:2013tza}.  Various nonrelativistic enhancements in dark matter annihilation and elastic scattering, invoked to solve various intriguing anomalies, can be related and attributed to the presence of an S-wave resonance very near to the scattering threshold of two dark matter particles.  If the resonance is  sufficiently near the scattering threshold, there is a region of energy in which the cross section comes close to saturating the unitarity bound and a single complex parameter, the S-wave scattering length, governs all the lower-energy behavior of the dark matter, i.e., the elastic and inelastic scattering cross section of two dark matter particles and the binding energy and lifetime of the resonance.  If the resonance is below the threshold, it is a bound state of the two dark matter particles.  If the dark matter particles have no annihilation channel, then the scattering length is real, the bound state is stable and a single real parameter governs the elastic scattering and the binding energy.  More generally, the scattering length also governs the low-energy few-body physics with more than two particles, such as loosely bound states consisting of three or more particles and the elastic scattering or the breakup scattering of these bound states\,\cite{Braaten:2004rn}.  These illustrate the principle of universality which we define in the next section.

Given the recent excitement about self-interacting dark matter, one can try to apply the new ideas mentioned in\,\cite{Braaten:2004rn,Braaten:2013tza} to other respects of dark matter physics.  Interactions between dark matter particles that come close to saturating the S-wave unitarity bound can naturally produce weakly bound states.  For example, a two-body bound state requires only that the scattering length be positive.  The binding energies and the low energy scattering properties of the weakly bound states are essentially determined by the same parameter, the scattering length, that governs the scattering of the individual particles.  Thus these bound states form a well-motivated and highly-constrained possibility for internal structure of dark matter.  It is intriguing to ask whether these bound states can have observable effects in searches for dark matter.  In this work, we point out that 2-body bound states provide novel features in the nuclear recoil energy spectrum in direct detection  experiments and therefore a smoking gun signature for internal structure in dark matter.

We assume that the self-interactions between dark matter particles are strong at low energies in the sense that they come close to saturating the S-wave unitarity bounds.  We also assume that the S-wave scattering length is positive, so that two dark matter particles form a weakly bound state.  (From here on, whenever we use the word ``particle", we will be referring to a single dark matter particle, which we will think of as a point particle; a bound state of dark matter particles will not be called a ``particle").  We assume the bound state is stable, so it can  act as a nonnegligible part of the dark matter of the Universe.  We assume that this bound state survives the cosmic evolution and the infrequent collisions with other particles.  These assumptions are not drastic: the deuteron is a weakly  bound state of the proton and the neutron, and we know from the very successful theory of Big Bang Nucleosynthesis, that it can survive from the very early Universe.  To be concise, we call this bound state of two dark matter particles ``darkonium".  Indeed, much of our formalism about the bound state can be identified as a dark copy of the deuteron.

We study the effect of this bound state in dark matter direct detection experiments.  Dark matter direct detection probes the elastic scattering of dark matter particles from a nucleus at relatively low energies\,\cite{Peter:2013aha,Freese:2012xd,Kurylov:2003ra,Fan:2010gt}.  If this energy scale is in the low-energy region where elastic self-scattering of the particles is governed by the scattering length, then the scattering of the bound state from the nucleus is also governed by the scattering length.  The scattering of this bound state with the target nucleus in a dark matter direct detection experiment will give a different nuclear recoil energy spectrum than the scattering of a dark matter particle.  This can be understood partly as the effect of the extended structure of the incoming bound state, which will  imprint a form factor on the recoil energy spectrum of the target nucleus, and partly due to the possibility of the breakup of the bound state by the scattering.  We do a complete calculation in this framework and find a new nuclear recoil energy spectrum.  If, in the future, such a structure is seen in the nuclear recoil energy spectrum, this will be a smoking gun signature for the internal structure for dark matter.

In Sec.\,\ref{sec:dark matter model}, we describe some of the universal properties of dark matter particles with a large scattering length.  In Sec.\,\ref{sec:derivation of nuclear recoil spectrum}, we present the expressions for the nuclear recoil energy spectrum due to an incident dark matter particle and an incident darkonium.  Sec.\,\ref{sec:nuclear recoil spectrum} gives some examples of the nuclear recoil energy spectrum for various nuclei that can be observed in dark matter direct detection experiments, comparing the spectrum from an incident flux of darkonium with that from an incident flux of dark matter particles.  We conclude in Sec.\,\ref{sec:conclusion}.  The details of the derivation of the nuclear recoil energy spectrum are given in the Appendix.

%%%%%%%%%%%%%%%%%%%%%%%%%%%%%%%%%%%%%%%%%%%%%%%%%%%
\section{Dark matter particles with large scattering lengths}
\label{sec:dark matter model}
%%%%%%%%%%%%%%%%%%%%%%%%%%%%%%%%%%%%%%%%%%%%%%%%%%%

The strong self-interaction cross sections at nonrelativistic velocities that are required to solve the small scale structure problems can motivate us to study other nonrelativistic systems in physics.  Due to the crucial availability of experimental data, the knowledge gained in these different systems might be extremely valuable in trying to understand the unknown properties of dark matter.

The success of the $\Lambda$CDM model implies that dark matter must have weak self-interactions at relativistic velocities, but it could have strong self-interactions at sufficiently small velocities.  In general, the strength of self-interactions is limited by the unitarity bounds of quantum mechanics.  We make the predictive assumption that the self-interactions of dark matter particles come close to saturating the S-wave unitarity bound in some velocity range.  We will refer to this velocity range as the {\it scaling region}.  In the scaling region, the scattering cross section for two dark matter particles have a power-law dependence on their relative velocity $v$.  For example, the elastic cross section is proportional to $1/v^2$.  At lower velocities, the cross sections are completely determined by a single parameter: the S-wave scattering length, which we denote by $a$\,\cite{Braaten:2013tza,Braaten:2004rn}.  This single parameter also controls other aspects of the low-energy few-body physics of the dark matter particles.  This is what makes the assumption so predictive.

A scaling region requires a resonance with an S-wave coupling to two dark matter particles that is very near their scattering threshold.  Such a resonance requires a fine-tuning.  The conditions for the fine tuning are most easily expressed in terms of the S-wave scattering length.  If there are dark matter annihilation channels, $a$ is complex with a small negative imaginary part.  We denote the range of the interaction between the dark matter particles by $r_0$.  The condition for the existence of a scaling region is that the scattering length must be large compared to the range: $|a| \gg r_0$.  The resonance could arise from interactions between the dark matter particles whose strength is tuned to near the critical value for there to be a bound state exactly at the threshold.  If such an interaction arises from the exchange of a particle of mass $m_y$ in the $t$-channel, the range is $r_0 \sim 1/m_y$.  The resonance could also be due to an elementary particle whose mass is very close to twice the mass of the dark matter particles and which has an S-wave coupling to the dark matter particles in the $s$-channel.  If the elementary particle has a mass $m_R$ and the tree-level cross section is $4\pi \alpha_R^2 \, m_R^2/|s - m_R^2|^2$, the relevant range of interactions is $r_0 = 1/(\alpha_R m_R)$.  Dark matter properties that are determined only by the S-wave scattering length are known as universal properties.  {\it Universality} in this context refers to the fact that systems with large scattering lengths have identical low-energy properties, independent of the underlying microphysics, if the variables are scaled by the appropriate factors of $|a|$.  The properties depend on the sign of $a$.  If $a$ is complex, they also depend on the ratio Im($a$)/Re($a$).

In the {\it universal region} defined by energies in and below the scaling region, the elastic scattering cross section and annihilation cross section for identical bosons can be written as~\cite{Braaten:2013tza}
\begin{eqnarray}
\sigma_{\rm el} = \dfrac{8 \pi}{\left | - i k - \gamma  \right |^2} \, ,
\label{eq:universal elastic cross section}
\end{eqnarray}
and 
\begin{eqnarray}
\sigma_{\rm ann} = \dfrac{8 \pi \, {\rm Im}(\gamma)}{k \left | - i k - \gamma  \right |^2} \, ,
\label{eq:universal annihilation cross section}
\end{eqnarray}
where $k$ is the relative momentum and $\gamma = 1/a$ is the inverse scattering length.  The relative momentum can be expressed as $k = \frac 12 \, m v$, where $v$ is the magnitude of the difference between the velocities of the two dark matter particles and $m$ is the mass of a dark matter particle.  If the two particles are distinguishable or if they are different spin states of identical fermions, we have to multiply the above equations by a factor of $\frac 12$.  In the above expressions, the $- ik$ term describes rescattering of the dark matter particles, which is an important effect if a resonance is sufficiently near the threshold~\cite{Braaten:2007nq}.  This term is proportional to the elastic width referred to in some previous literature~\cite{1937PhRv...51..450B,Landau:1991,MarchRussell:2008tu}. 

In the universal region, the properties of the resonance are also determined by the scattering length $a$\,\cite{Braaten:2013tza}.  In particular, if ${\rm Re}\, \gamma > 0$, the resonance is a bound state of the two  dark matter particles, with a finite lifetime.  The binding energy of the resonance is 
\begin{eqnarray}
E_B = \dfrac{({\rm Re}\, \gamma )^2 - ({\rm Im}\, \gamma )^2}{m}\, ,
\label{eq:binding energy wimpnoium}
\end{eqnarray}
and the lifetime of the bound state is 
\begin{eqnarray}
\Gamma_{\rm darkonium} =\dfrac{4 \,({\rm Re}\, \gamma ) \, ({\rm Im}\, \gamma )}{m}\, .
\label{eq:decay rate wimpnoium}
\end{eqnarray}
The Schrodinger wave function of the bound state is
\begin{eqnarray}
\psi (r) = \sqrt{\dfrac{{\rm Re} \, \gamma}{2 \pi}} \, e^{- \gamma \, r}/r \, .
\end{eqnarray}
Thus the bound state has a spatial extent 1/(${\rm Re} \, \gamma$) that is much larger than the range of the interactions between the dark matter particles.  The large separation of the two dark matter particles in the bound state is a remarkable phenomenon.  It is particularly remarkable in the case of a resonance that arises from an elementary particle whose mass is very close to twice that of the dark matter particle.

There are many examples in Nature of particles with large scattering lengths\,\cite{Braaten:2004rn}.  In nuclear physics, the classic example is the neutron, which has a large negative scattering length.  Neutrons and protons have a relatively large positive scattering length in the isospin-0 channel.  The associated bound state is the deuteron.  In atomic physics, the spin-triplet state of the tritium atom $^3$H has a large negative scattering length.  The helium atom $^4$He has a large positive scattering length.  The associated bound state is the diatomic $^4$He molecule, which has a very tiny binding energy of about 10$^{-7}$ eV.  In high energy physics, the charm mesons $D^0$ and $D^{*0}$ have a large positive scattering length in the even charge conjugation channel\,\cite{Braaten:2003he}.  The associated bound state is called the X(3872).  These are all examples in which Nature has produced an accidental fine-tuning of an S-wave resonance to near the appropriate threshold.  It is possible that Nature has also provided an analogous fine-tuning for dark matter.  

In the field of cold atom physics, the scattering length for atoms can be controlled by the experimenter.  By tuning a magnetic field to near a Feshbach resonance where the energy of the diatomic molecule is at the scattering threshold for a pair of atoms, the scattering length can be made arbitrarily large (or small)\,\cite{2010RvMP...82.1225C}.  This has allowed detailed experimental studies of the few-body physics and many-body physics of particles with large scattering lengths.  These experiments may be directly applicable to dark matter if it consists of particles with a large scattering length.  

In our case, we wish to consider a bound state that has a very long lifetime.  This amounts to taking the limit ${\rm Im}\, \gamma \rightarrow$ 0.  From Eqn.\,(\ref{eq:universal annihilation cross section}), it is clear that this requires a vanishing annihilation cross section.  A vanishing annihilation cross section is most easily accommodated by dark matter sector that is asymmetric, just like the visible sector. 

In the limit of ${\rm Im}\, \gamma \rightarrow$ 0, the self-interaction cross section in Eqn.\,(\ref{eq:universal elastic cross section}) reduces to
\begin{eqnarray}
\sigma_{\rm el} = \dfrac{8 \pi}{{\gamma}^2 + k^2}\, .
\label{eq:real gamma sigma elastic}
\end{eqnarray}   
The binding energy in Eqn.\,(\ref{eq:binding energy wimpnoium}) reduces to
\begin{eqnarray}
E_{B} = \dfrac{\gamma^2}{m} \, .
\label{eq:real gamma binding energy}
\end{eqnarray}
The elastic scattering cross section and the binding energy are determined by the single real parameter $\gamma$.  This parameter also determines other aspects of the low-energy few-body physics of the dark matter particles.  In particular, it determines up to an overall normalization factor the scattering of darkonium with small momentum transfer from a nucleus in dark matter direct detection experiments.  The finite size of the darkonium may produce a tell-tale signature in the recoil energy spectrum of the target nucleus.  The breakup of the darkonium into two dark matter particles from scattering off the nucleus could also provide a signature.  These provide the main motivation for calculating the recoil energy spectrum of the target nucleus for an incident darkonium.  If it is possible to infer that a component of the dark matter is a universal bound state with the inverse scattering length $\gamma$, then using Eqn.\,(\ref{eq:real gamma sigma elastic}), one can easily infer the dark matter self-interaction cross section.

Our basic premise is that there is a scaling region of the relative velocity $v$ in which dark matter particles come close to saturating the S-wave unitarity bound:  $\sigma_{\rm el} \approx 32 \pi/m^2 \, v^2$.  We should therefore ask whether such large cross sections are compatible with the known properties of dark matter.  Since the unitarity bound is proportional to $1/m^2$, an upper bound on $\sigma_{\rm el}$ from astrophysics can always be accommodated by making the dark matter mass sufficiently large.  One such upper bound comes from the Bullet Cluster, for which the observed mass distribution from gravitational lensing sets an upper bound on the elastic cross-section divided by the mass of the dark matter particle: $\sigma_{\rm el}/m \lesssim 1$ cm$^2$ g$^{-1}$ at the estimated collision velocity of $v \sim$ 1000 km s$^{-1}$.  This is consistent with the unitarity bound being saturated at that velocity if $m \lesssim$ 12 GeV.  A larger mass would require this velocity to be above the scaling region.  Another possible constraint comes from the small scale structure problems in $\Lambda$CDM.  The absence of a cusp in the dark matter distribution of dwarf galaxies can be explained by self-interactions of dark matter particles whose order of magnitude is $\sigma_{\rm el}/m \approx$ 1 cm$^2$ g$^{-1}$ at the typical velocity of $v \approx$ 10 km s$^{-1}$.  This is consistent with the unitarity bound being saturated at that velocity if $m \approx$ 270 GeV.  A smaller mass can be accommodated if the elastic cross section in Eqn.\,(\ref{eq:real gamma sigma elastic}) is already approaching its low energy limit $\sigma_{\rm el} = 8 \pi/\gamma^2$ at that velocity.  Thus cross sections with a scaling region in which the unitarity bound is nearly saturated are compatible with the known properties of dark matter with mass in the range relevant to current experiments.  

The scattering of darkonium is also determined by the inverse scattering length $1/\gamma$.  When two darkonia collide, they can scatter elastically or inelastically.  If the scattering is inelastic, there are several possibilities for the final state.  It can consist of 4 individual dark matter particles, or a darkonium plus 2 individual dark matter particles, or a bound state of 3 dark matter particles plus an individual dark matter particle.  If some light particle (such as a dark photon) can be radiated in the collision, the final state can also be a bound state comprised of 4 dark matter particles.  The possibility of forming bound states comprised of 3 or more dark matter particles can be avoided by imposing certain symmetries, as in the case of a spin$-\frac12$ dark matter particle.  The formation of bound states can also be avoided through decay instabilities.  For example, in the visible world, only nuclei with specific proton and neutron numbers are stable.
 
The calculation of the elastic darkonium self-scattering cross section is a non-trivial 4-body problem.  Generically, the low-energy elastic cross section is the same order of magnitude as that for the elastic scattering of the dark matter particles in Eqn.\,(\ref{eq:real gamma sigma elastic}), which is of order $1/\gamma^2$.  For example, if the constituents of the darkonium are the two spin states of a spin-$\frac 12$ fermion, the darkonium scattering length is $0.6/\gamma$\,\cite{Petrov:2004zz}.  If the constituents of the darkonium are a spin-0 boson, the darkonium scattering length is $1/\gamma$ multiplied by a log-periodic function of $\gamma$ that has the same value when $\gamma$ is changed by a multiplicative factor of 22.7\,\cite{Braaten:2004rn}.  For most values of $\gamma$, the darkonium scattering length is between $-3/ \gamma$ and $+3/ \gamma$, but it is much larger near the critical values of $\gamma$ for which there is a 4-boson bound state at the 2-darkonium threshold\,\cite{Deltuva:2011ur}.  At high energies, the total darkonium self-scattering cross section is also the same order of magnitude as that for the dark matter particles, which is of order $1/k^2$.  However, the elastic darkonium self-scattering cross section is much small, scaling as $\gamma^4/k^6$.  The suppression factor of $(\gamma/k)^4$ arises because the momentum transfer must be transmitted to both constituents of both the darkonia.

To measure the probability of darkonium breakup, we calculate the mean free path, where we take $\sigma_{el}/m$ = 1 cm$^2$ g$^{-1}$.  The calculation in this paragraph is only an order of magnitude estimate to get a sense of scales involved.  In general, whether a darkonium will survive can only be addressed in a detailed $N$-body simulation.  If the background dark matter density is cosmological (i.e., 1.26 $\times$10$^{-6}$ GeV cm$^{-3}$), then the mean free path of the darkonium is approximately 150 Gpc.  This result is independent of the dark matter particle mass as a higher mass means stronger self-interaction and it also implies lower number of dark matter particles.  If the background dark matter density is 0.1 GeV cm$^{-3}$, then the mean free path is approximately 2 Mpc.  For a higher dark matter density, the mean free path will be lower and hence the chance of darkonium breaking up will be higher.    From the above arguments, it is clear that unless the darkonium passes though a region of fairly high dark matter density, the survival probability will be quite high.

%%%%%%%%%%%%%%%%%%%%%%%%%%%%%%%%%%%%%%%%%%%%%%%%%%%
\section{Nuclear recoil energy spectrum}
\label{sec:derivation of nuclear recoil spectrum}
%%%%%%%%%%%%%%%%%%%%%%%%%%%%%%%%%%%%%%%%%%%%%%%%%%%

In this section, we present the nuclear recoil energy spectrum that is measured in a dark matter direct detection experiment.  We will first give the nuclear recoil energy spectrum of a dark matter particle scattering off a nucleus, followed by the nuclear recoil energy spectrum of a bound state of two dark matter particles (darkonium) scattering off a nucleus.  For darkonium scattering off a nucleus, there are two possible final states: (a) the darkonium can remain bound after the scattering, and (b) the darkonium can be broken apart due to the scattering with the nucleus.  The details of the derivation of the nuclear recoil energy distribution are given in the Appendix.

We assume for simplicity that the two constituents of darkonium have equal mass $m$.  They can be identical bosons or different spin states of a spin-$\frac12$ particle or distinguishable particles.  The mass of the darkonium can be approximated by 2$m$.  We denote the mass of the target nucleus by $m_A$.  The magnitude of the momentum transferred to the nucleus is denoted by $q$.  The nuclear recoil energy is $E_{\rm nr} = q^2/2 m_A$.  We also assume for simplicity that the two constituents of darkonium scatter from the nucleus with the same amplitude $G_A(q)$.

\subsection{Dark matter particle scattering off the nucleus}

\begin{figure}[!h]
\centering
\includegraphics[angle=0.0,width=0.45\textwidth]{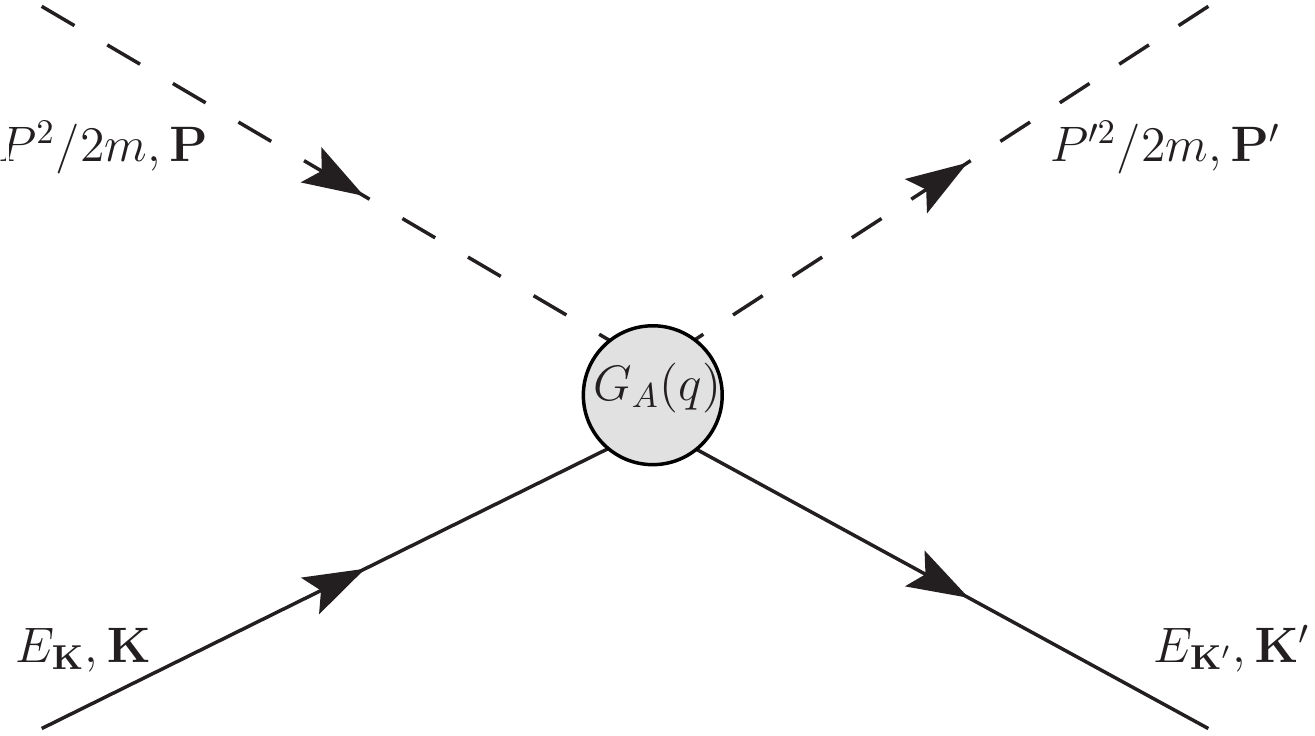}
\caption{Feynman diagram for a dark matter particle scattering off a target nucleus.  The incoming and outgoing dark matter particles have momenta ${\bm P}$ and ${\bm P'}$ and are shown by single dashed lines.  The incoming and outgoing nucleus have momenta ${\bm K}$ and ${\bm K}'$ and are shown by solid lines.  Energies and momenta are denoted by normal font and bold letters respectively.  The vertex for the effective interaction of a single dark matter particle with the nucleus is represented by the grey blob.}
\label{fig:Feynman diagram particle scattering}
\end{figure}

In this section, we give the recoil energy spectrum of the scattered nucleus due to scattering with a dark matter particle.  The Feynman diagram is shown in Fig.\,\ref{fig:Feynman diagram particle scattering}.
The differential scattering rate of one dark matter particle, with velocity $v$, off a target nucleus is
\begin{eqnarray}
\left(\dfrac{d (\sigma v)}{dE_{\rm nr}}\right)_{{A + 1}} = \dfrac{m_A}{v}\,|G_{A} (q)|^2 \dfrac{1}{2 \pi } \, \Theta \left(v - q/2 \mu \right) \, ,
\label{eq:differential scattering rate wimp}
\end{eqnarray}
where $G_A(q)$ is the vertex factor for the effective interaction between the dark matter particle and the nucleus.  There is a minimum velocity of the dark matter particle necessary to produce a recoil of momentum $q$: $v > q/2 \mu$, where $\mu$ is the reduced mass of the dark matter particle and the nucleus.

Comparing the expression in Eqn.\,(\ref{eq:differential scattering rate wimp}) with the standard expression in the literature for the case of a spin-independent cross section $\sigma_{\rm SI}$ between the dark matter particle and the nucleon, we find that 
\begin{eqnarray}
|G_{A}(q)|^2 = \dfrac{\pi \, \sigma_{\rm SI} A^2 F_N^2(q)}{\mu_n^2} \, ,
\label{eq:effective interaction} 
\end{eqnarray}
where $A$ is the mass number of the target nucleus, $\mu_n$ is the reduced mass of the dark matter particle and the nucleon, and $F_N(q)$ is the nuclear form factor.

\subsection{Bound state elastic scattering off the nucleus}

\begin{figure}[!h]
\centering
\includegraphics[angle=0.0,width=0.45\textwidth]{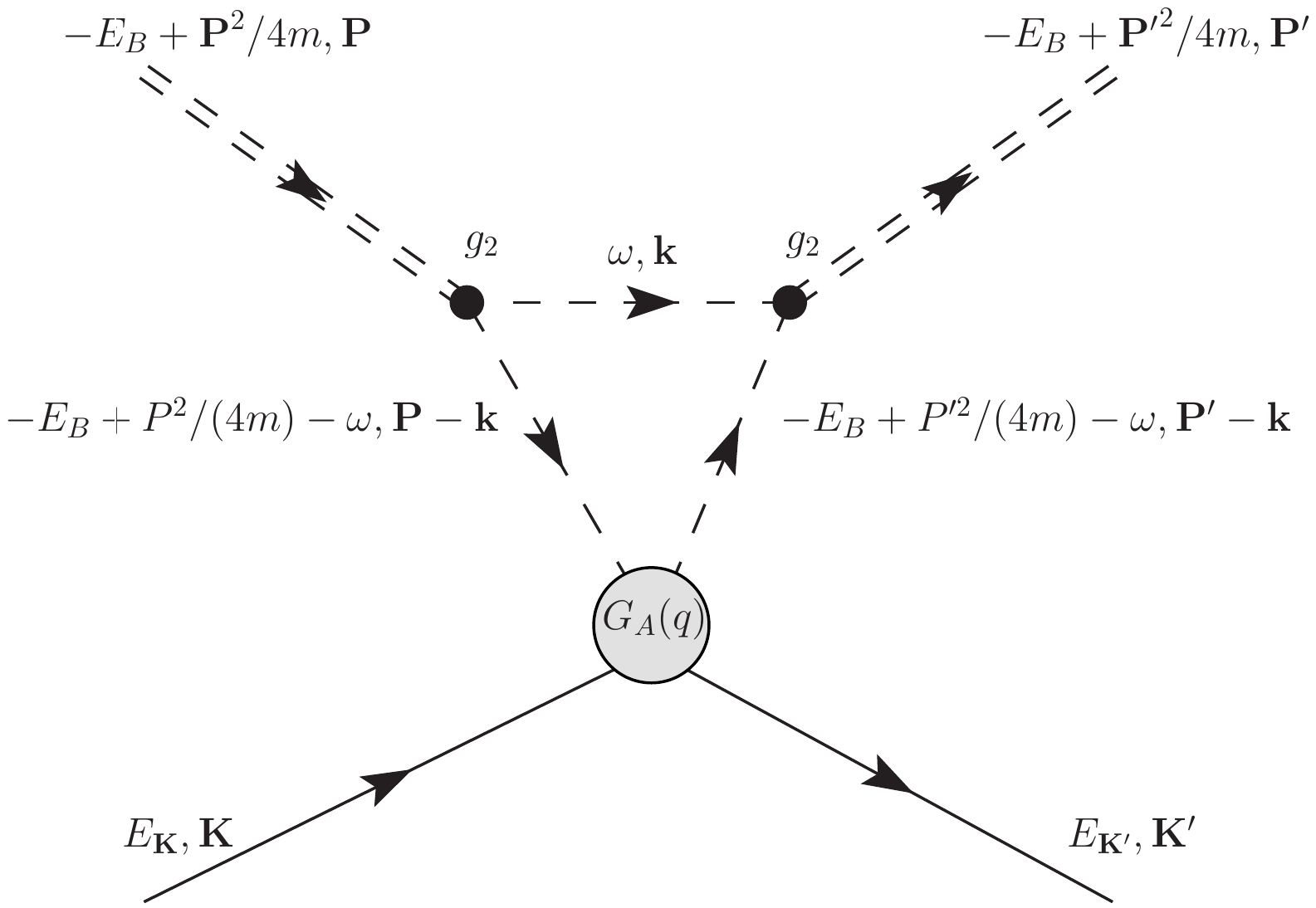}
\caption{Feynman diagram for a darkonium scattering off a target nucleus.  The incoming and outgoing darkonium have momenta ${\bm P}$ and ${\bm P'}$ and are shown by the double dashed lines.  All other notations are the same as in Fig.\,\ref{fig:Feynman diagram particle scattering}.}
\label{fig:Feynman diagram darkonium scattering}
\end{figure}

In this section, we give the recoil energy spectrum of a scattered nucleus due to a darkonium elastically scattering off the target nucleus.  The Feynman diagram for this process is shown in Fig.\,\ref{fig:Feynman diagram darkonium scattering}.  One of the constituents of the darkonium scatters from the nucleus and subsequently recombines with the other constituent to form darkonium again.

\begin{figure*}[!tbhp]
\centering
\includegraphics[angle=0.0,width=0.45\textwidth]{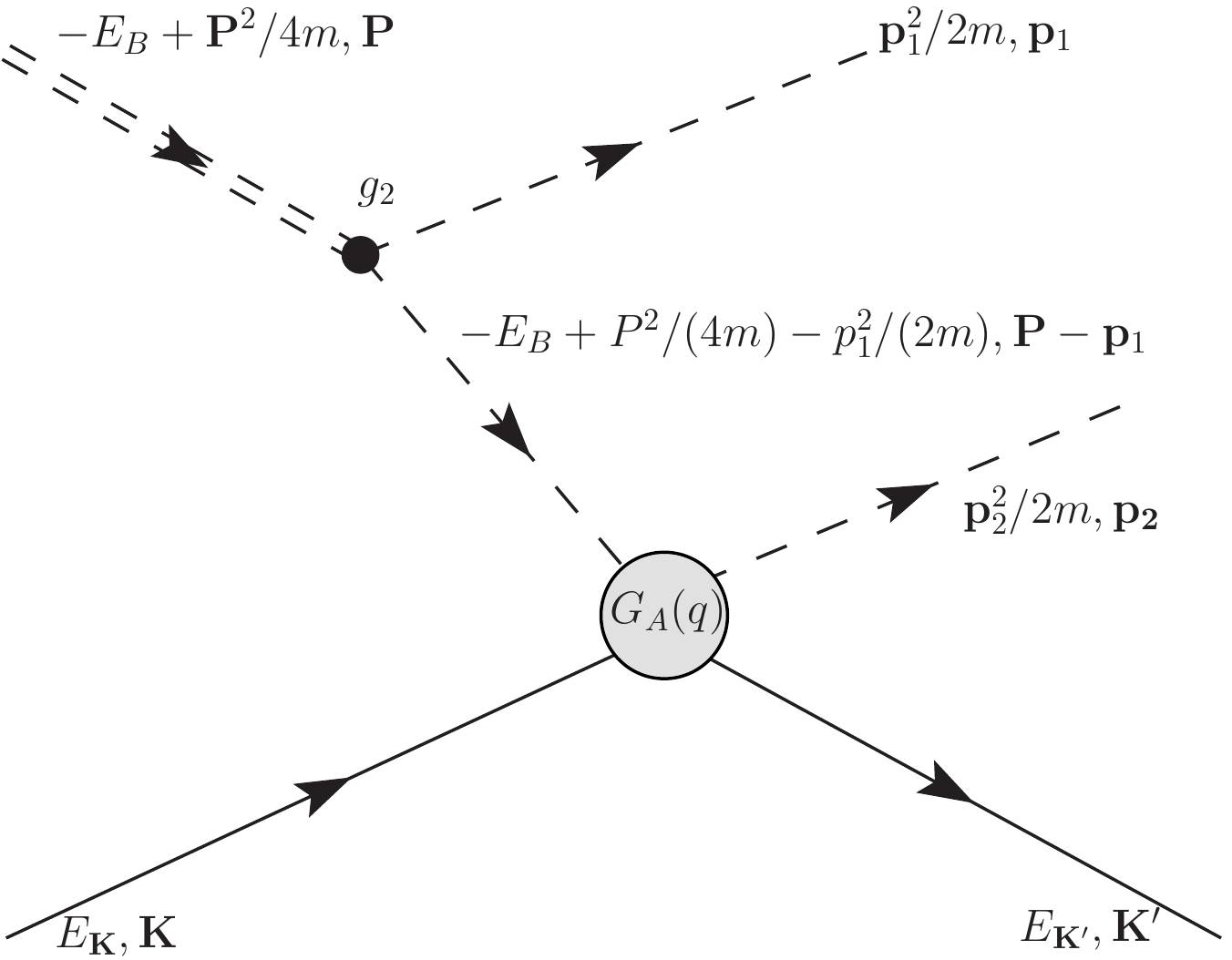}
\includegraphics[angle=0.0,width=0.5\textwidth]{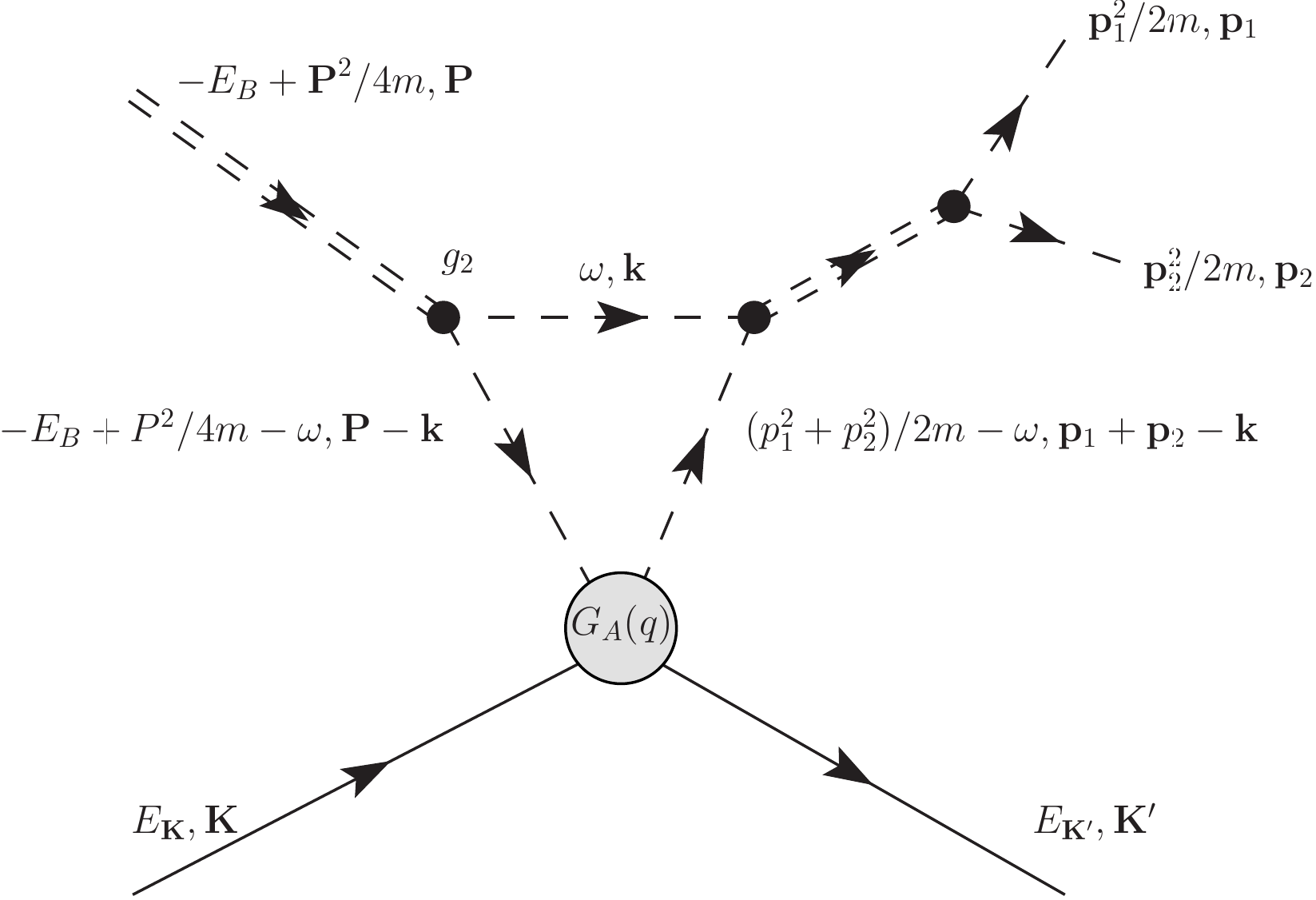}
\caption{Feynman diagrams for a darkonium breakup from scattering with a target nucleus.  The momenta of the outgoing dark matter particles are ${\bm p}_1$ and ${\bm p}_2$.  There is one more diagram which is identical to the diagram on the left but with ${\bm p}_1$ and ${\bm p}_2$ interchanged.  All other notations are the same as in Figs.\,\ref{fig:Feynman diagram particle scattering} and \ref{fig:Feynman diagram darkonium scattering}.}
\label{fig:Feynman diagram darkonium breakup}
\end{figure*}

The Feynman diagram in Fig.\,\ref{fig:Feynman diagram darkonium scattering} is calculated in the Appendix.  The differential rate of one darkonium with a velocity $v$ to scatter elastically off a target nucleus is
\begin{eqnarray}
\left(\dfrac{d (\sigma v)}{dE_{\rm nr}} \right)_{A+2} = \dfrac{m_A}{v}\,|G_{A}(q)|^2 \, \dfrac{2}{\pi } |F(q)|^2 \nonumber\\
\times  \Theta \left(v - q/2 \mu_2 \right) \, .
\label{eq:differential rate of darkonium scattering}
\end{eqnarray}
The form factor of the darkonium is given by
\begin{eqnarray}
F(q) = \dfrac{4 \gamma}{q} \, {\rm tan}^{-1} \left(\dfrac{q}{4 \gamma} \right) \, ,
\label{eq:darkonium form factor}
\end{eqnarray}
where $\gamma = 1/a$ is the inverse scattering length.  In the limit of small $q$, the form factor goes to 1.  In the limit of large $q/4 \gamma$, the form factor goes to $2 \pi \gamma/q$.  There is a minimum velocity of darkonium necessary to produce a nuclear recoil of momentum $q$: $v \geq q/2 \mu_2$, where $\mu_2$ is the reduced mass of the darkonium and the nucleus. 

The expression in Eqn.\,(\ref{eq:differential rate of darkonium scattering}) differs from the expression for a dark matter particle scattering off a nucleus in Eqn.\,(\ref{eq:differential scattering rate wimp}) by the presence of the form factor, by a different argument of the theta function, which gives a minimum velocity required for the nuclear recoil momentum $q$, and by a factor of 4.  This factor of 4 (= $2^2$) can be understood as arising from the coherence effect of the darkonium which is composed of two dark matter particles.

\subsection{Bound state breakup from scattering off nucleus}

Here we give the nuclear recoil energy spectrum due to a darkonium break up from scattering off a nucleus.   The Feynman diagrams for this process are shown in Fig.\,\ref{fig:Feynman diagram darkonium breakup}.  In both of the diagrams in Fig.\,\ref{fig:Feynman diagram darkonium breakup}, one of the constituents of the darkonium scatters from the nucleus.  In the second diagram, the two constituents subsequently rescatter.  Because the interaction associated with a large scattering length is nonperturbative, this diagram must be included for consistency.  The diagrams are calculated in the Appendix.  The differential scattering rate for one darkonium to breakup after scattering with the target nucleus is
\begin{eqnarray}
&&\hspace{-1 cm}\left(\dfrac{d (\sigma v)}{dE_{\rm nr}} \right)_{A+1+1} = 128 \, \gamma \dfrac{m_A}{v} |G_A(q)|^2 \nonumber\\
&\times& \int \dfrac{d^3 {\bm r}}{(2 \pi)^3} \Bigg| \dfrac{1}{4 \gamma^2 + (2 {\bm r} - {\bm q})^2}
+ \dfrac{1}{4 \gamma^2 + (2 {\bm r} + {\bm q})^2} \nonumber\\
&& \hspace{1.5 cm}-\dfrac{i}{2q (\gamma + i r)} \, {\rm ln} \dfrac{4 r^2 + (2 \gamma - i q)^2}{4 \gamma^2 + (2 r - q)^2}\Bigg|^2 \nonumber\\
&\times& \Theta \left(v - \left(\dfrac{q}{2 \mu_2} + \dfrac{\gamma^2}{m q} \right) \right) \, .
\label{eq:darkonium breakup phase space simplified compact}
\end{eqnarray}

The integral over the angles of ${\bm r}$ can be calculated analytically to give a function of $r$ and $q$ that can be expressed in terms of logarithms.  The range of the subsequent integral over $r$ is $0 < r < R$, where $R$ depends on $v$, $q$ and $\gamma$:
\begin{eqnarray}
R^2 = m q \left(v - \left(\dfrac{q}{2 \mu_2} + \dfrac{\gamma^2}{m q} \right) \right) \, .
\label{eq:R}
\end{eqnarray}

The condition for validity of the recoil energy distribution in Eqn.\,(\ref{eq:darkonium breakup phase space simplified compact}) is $q/2 \ll 1/r_0$ and $R \ll 1/r_0$, where $r_0$ is the range of dark matter interactions.  The theta function in Eqn.\,(\ref{eq:darkonium breakup phase space simplified compact}) implies that the breakup of darkonium is possible only if its velocity $v$ in the laboratory frame exceeds $q/(2 \mu_2) + \gamma^2/(m q)$.   

%%%%%%%%%%%%%%%%%%%%%%%%%%%%%%%%%%%%%%%%%%%%%%%%%%%
\section{Recoil energy spectra off various nuclei}
\label{sec:nuclear recoil spectrum}
%%%%%%%%%%%%%%%%%%%%%%%%%%%%%%%%%%%%%%%%%%%%%%%%%%%

In this section, we will calculate some example nuclear recoil energy spectra for various target nuclei used in current dark matter direct detection experiments.  To cover the typical ranges of dark matter particle masses searched for in these experiments, we use two dark matter particle masses: 
\begin{itemize}
\item ``traditional" dark matter particles with a representative mass being $m$ = 100 GeV,  
\item light dark matter particles, with a representative mass being $m$ = 10 GeV.
\end{itemize}

\begin{figure*}[!thpb]
\centering
\includegraphics[angle=0.0,width=0.43\textwidth]{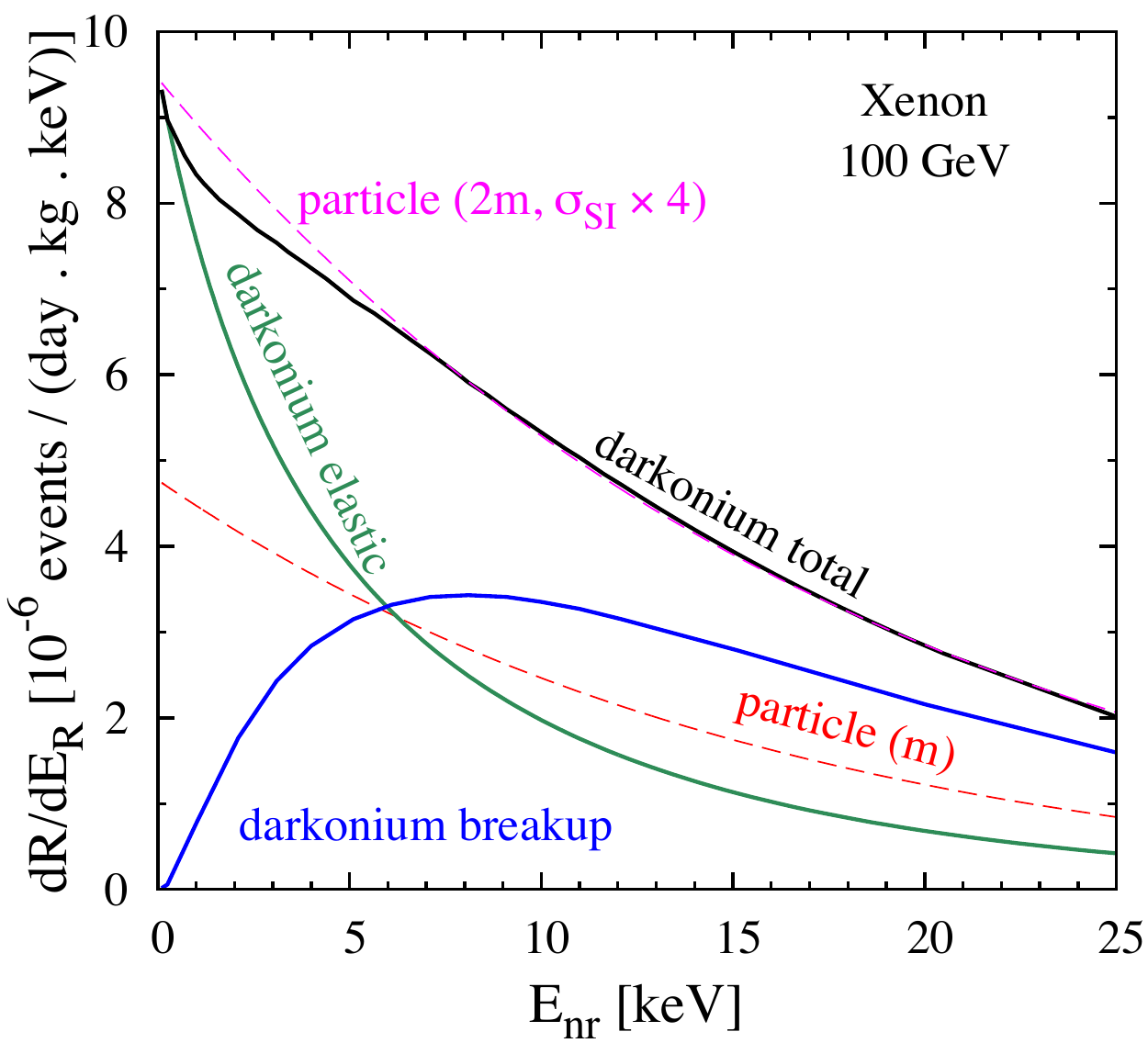}
\includegraphics[angle=0.0,width=0.43\textwidth]{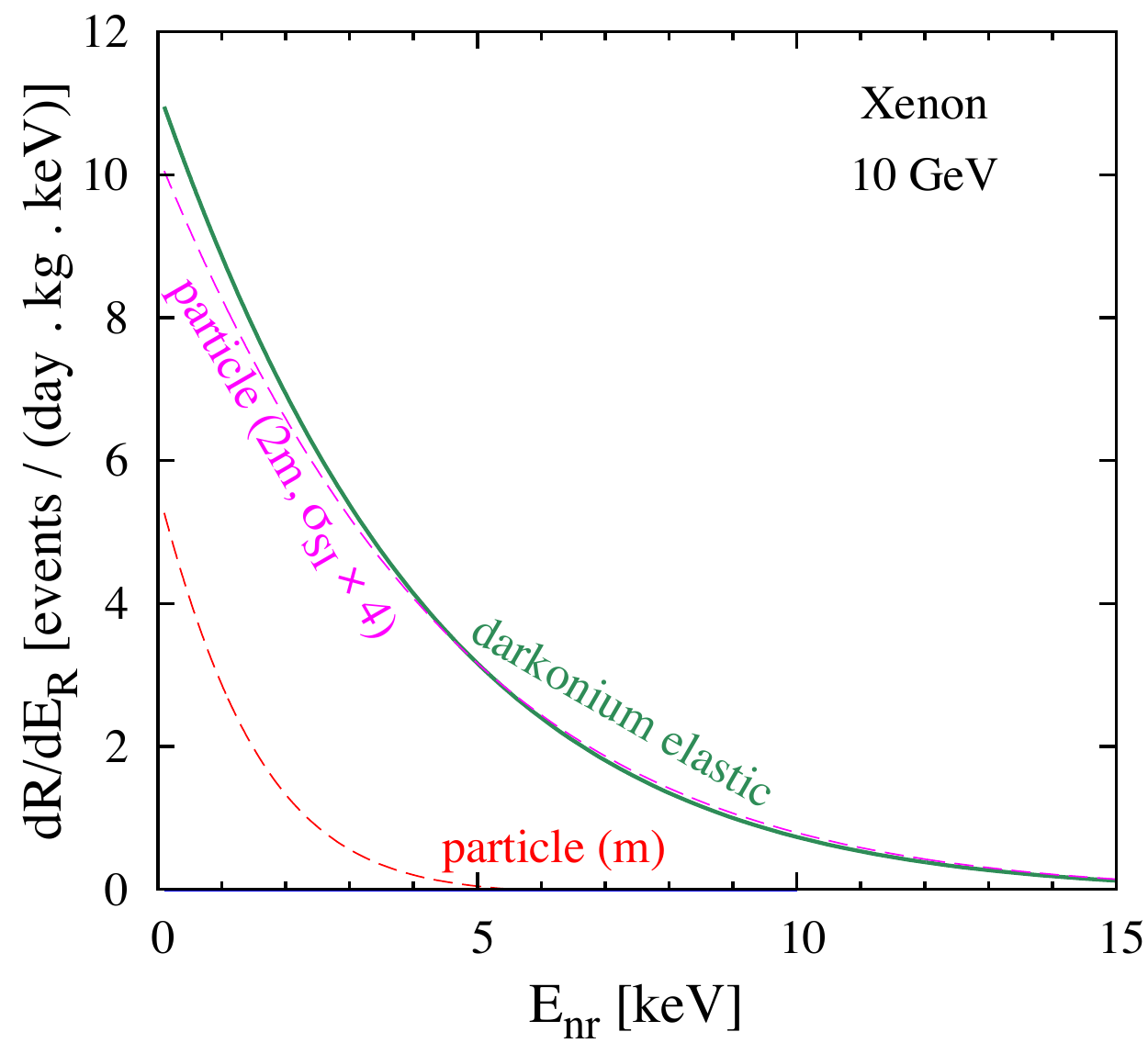}
\includegraphics[angle=0.0,width=0.43\textwidth]{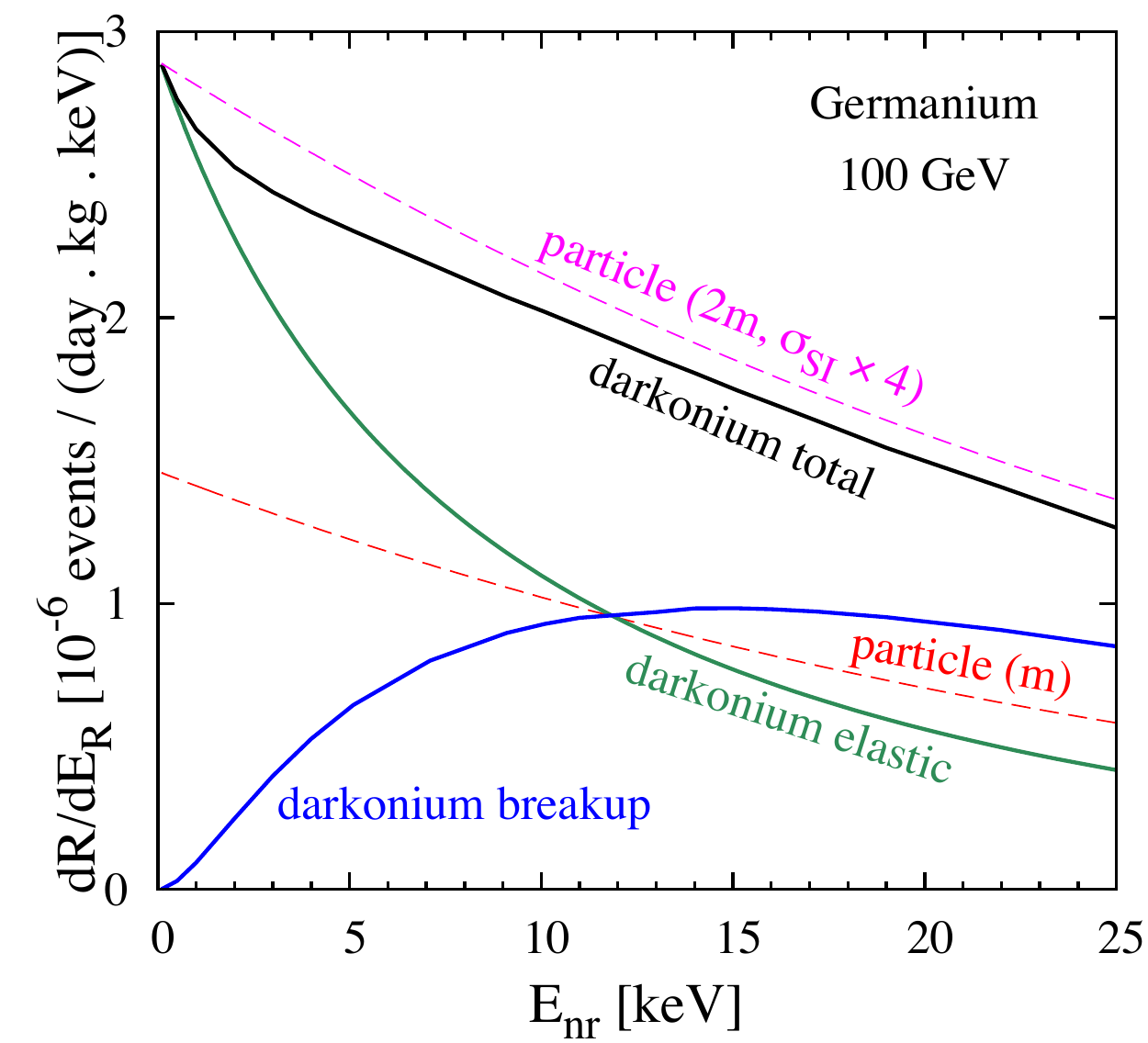}
\includegraphics[angle=0.0,width=0.43\textwidth]{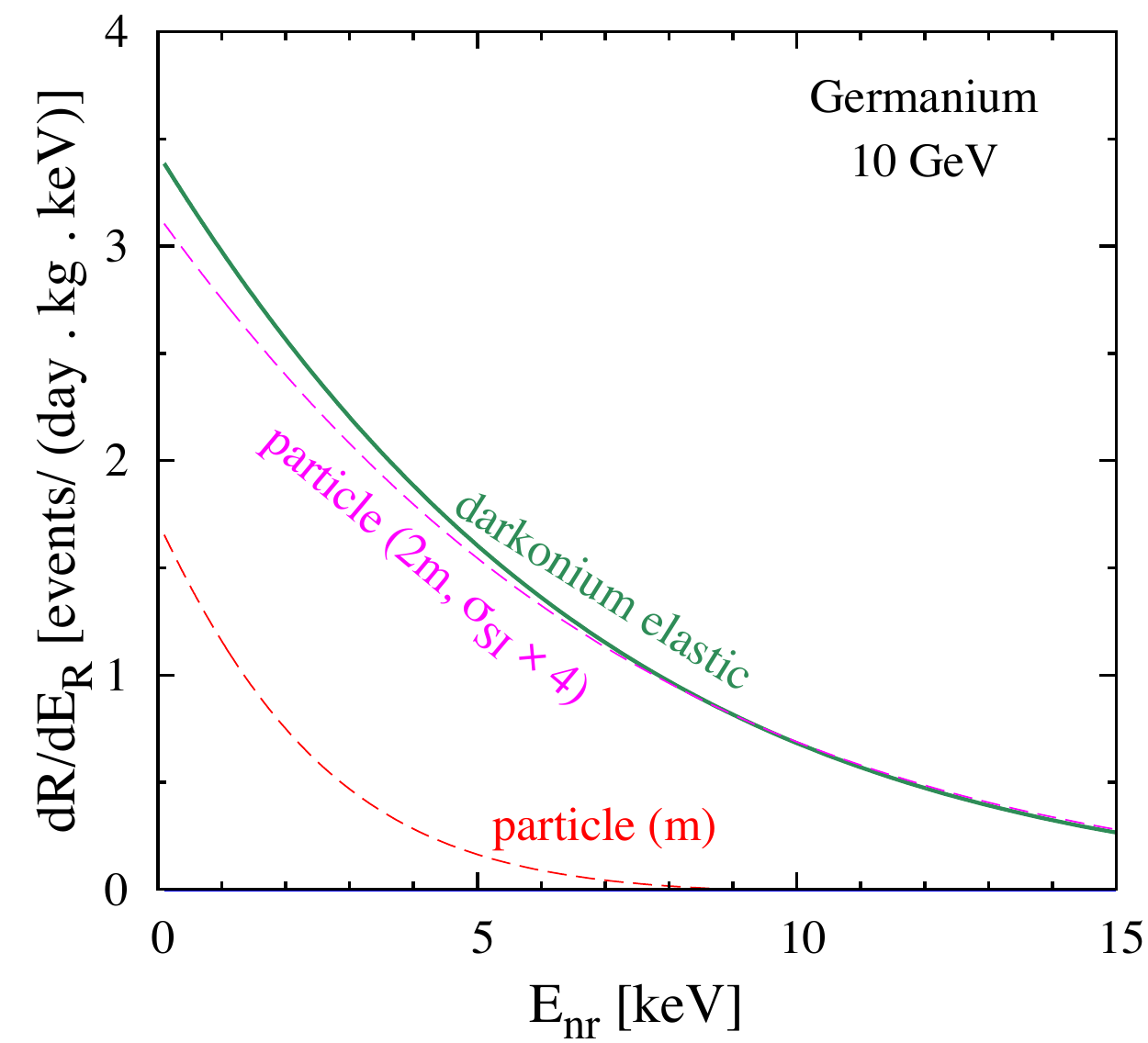}
\includegraphics[angle=0.0,width=0.43\textwidth]{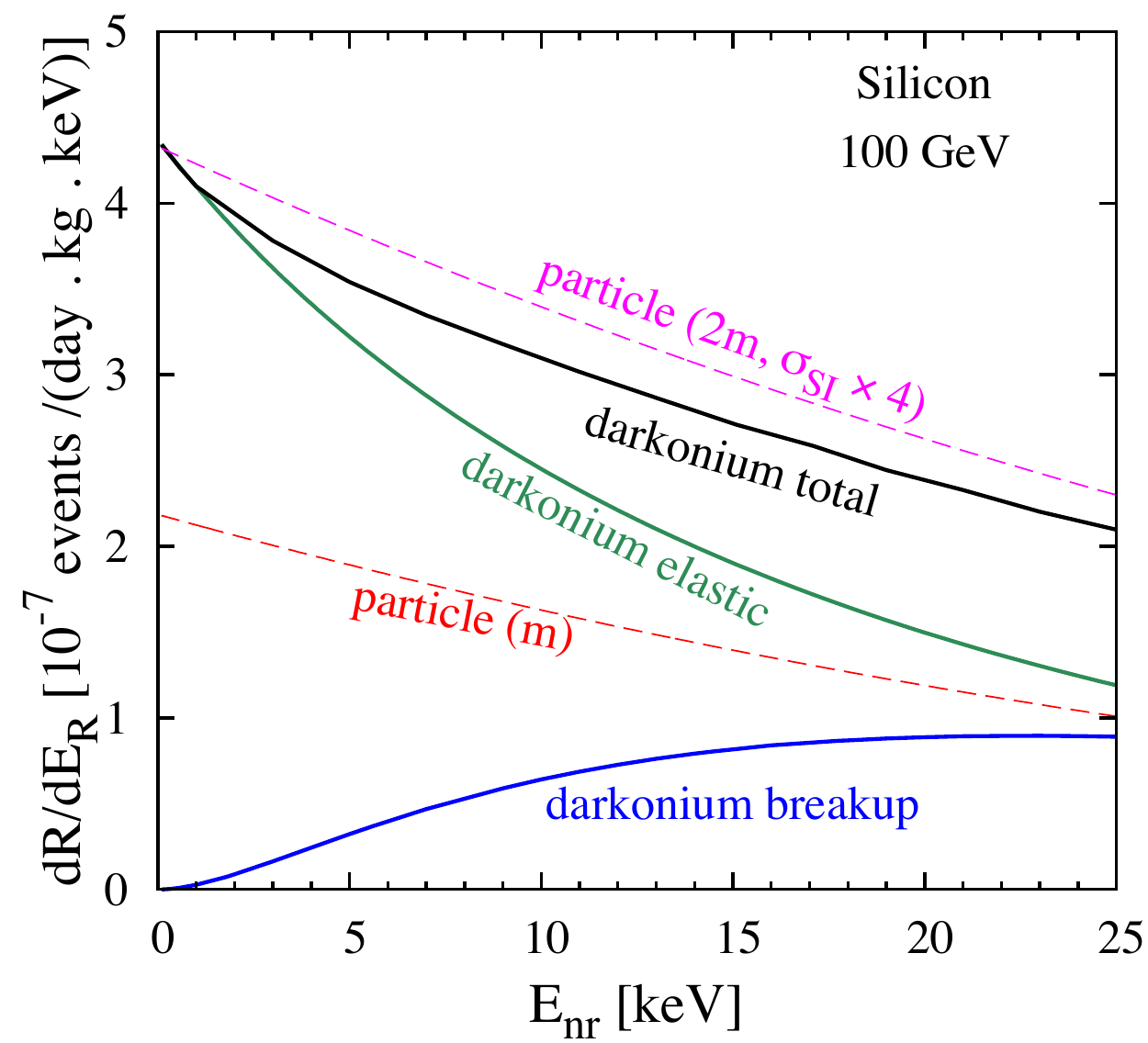}
\includegraphics[angle=0.0,width=0.43\textwidth]{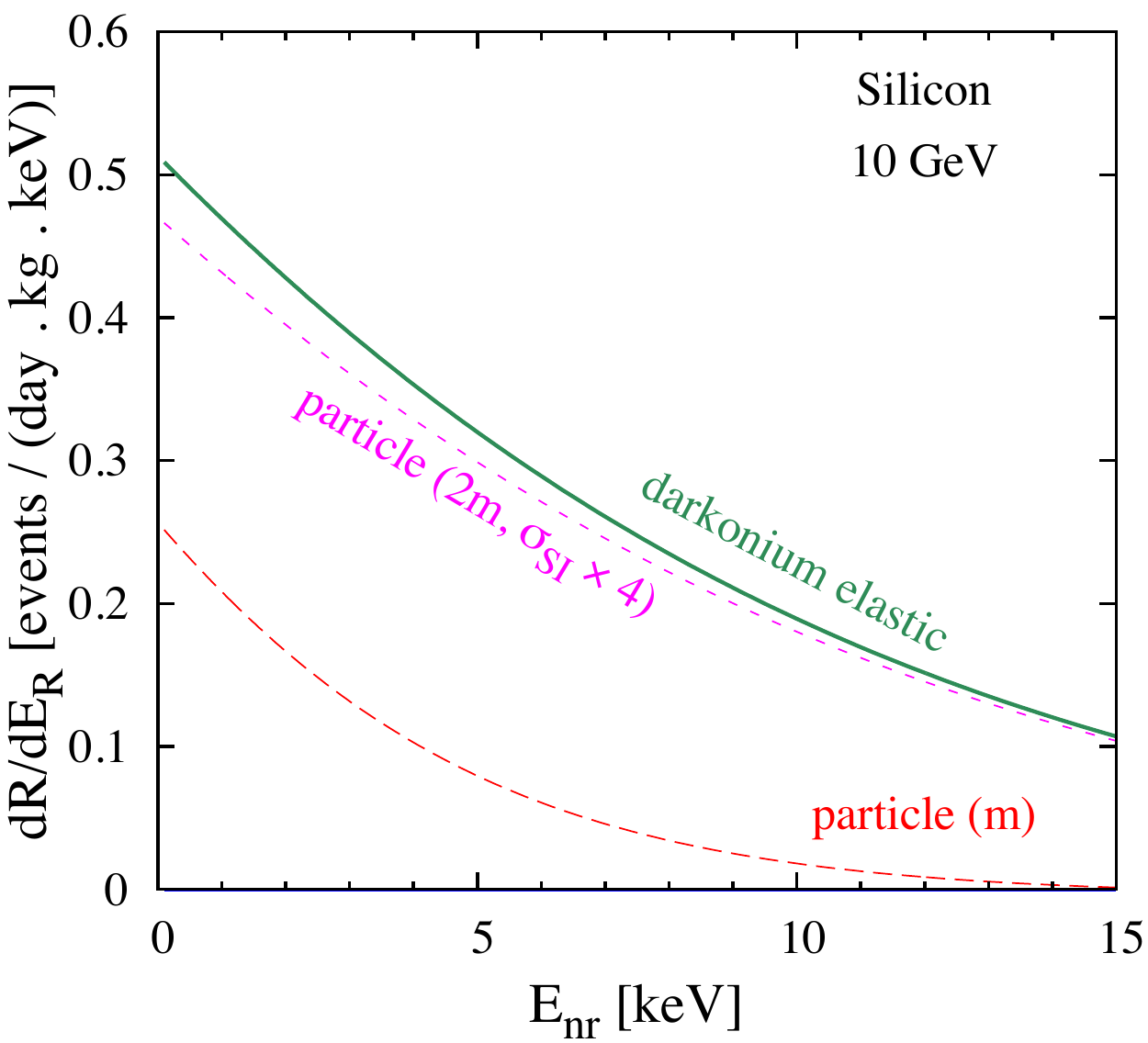}
\caption{The recoil energy spectra for dark matter particle (of mass $m$) scattering (red dashed),  darkonium elastic scattering (green solid), darkonium break up scattering (blue solid), and total darkonium scattering (black solid) with a target nucleus.  The element of the target nucleus and the mass of the dark matter particle are given in the top right hand corner of each plot.  For $m = 10$ GeV, the total darkonium scattering is the same as the elastic darkonium scattering.  We have determined $\gamma$ by taking the elastic scattering cross section per unit mass to be $\sigma_{\rm el}/m$ = 1 cm$^2$ g$^{-1}$ at $v = 10$ km s$^{-1}$.  Different values of $\sigma_{\rm SI}$ are used for the two masses. For comparison, the recoil energy spectrum is also shown for a dark matter particle with
mass $2m$ and $\sigma_{\rm SI}$ that is 4 times larger (dashed magenta line).}
\label{fig:Recoil spectra}
\end{figure*}

We will show the nuclear recoil spectra for three different nuclei: 
\begin{itemize}
\item Xenon, for which the atomic mass number $A$ ranges from 124 to 136, 
\item Germanium, for which $A$ ranges from 70 to 76,
\item Silicon, for which $A$ ranges from 28 to 30.
\end{itemize}
These span the range of nuclei that have good sensitivity for heavy dark matter and light dark matter candidates.  

We take the simplest case of an isospin-conserving, momentum-independent, spin-independent cross section between the dark matter particle and the nucleon to arrive at the expression for $G_{A}(q)$ in Eqn.\,(\ref{eq:effective interaction}).  We take the nuclear form factor $F_N(q)$ to be the Helm form factor\,\cite{Lewin:1995rx}.

The normalizations of our cross sections are determined by the spin-independent dark matter particle-nucleon cross section $\sigma_{\rm SI}$.  For the case of $m$ = 100 GeV, we choose $\sigma_{\rm SI}$ = 10$^{-46}$ cm$^2$, which is just beyond the present limit as presented by the XENON100 collaboration\,\cite{Aprile:2012nq} and the LUX collaboration\,\cite{Akerib:2013tjd}.  For the case of $m = 10$ GeV, we choose $\sigma_{\rm SI} = 10^{-41}$ cm$^2$, which is excluded by the recent XENON100 dataset\,\cite{Aprile:2012nq} and the LUX dataset\,\cite{Akerib:2013tjd}.  However, the present status of this region is controversial, as there are a number of anomalies which cannot be explained by known backgrounds but can be explained as due to dark matter scattering\,\cite{Angloher:2011uu,Aalseth:2012if,Agnese:2013rvf,Bernabei:2013xsa}.  These anomalies can be reconciled either by exotic physics or by improvements in experimental measurements.  These values of $\sigma_{\rm SI}$ are chosen only for illustration.  Other values of $\sigma_{\rm SI}$ would change the normalization of the recoil energy spectrum, keeping the shape unchanged.

Given the differential scattering rate of a single dark matter particle or darkonium scattering with the target nucleus, $(d(\sigma v)/dE_{\rm nr})_{\rm single}$, we can calculate the differential scattering rate (in units of events per unit time per unit target mass and per unit recoil energy) for an incident flux of dark matter as
\begin{eqnarray}
\left(\dfrac{dR}{dE_{\rm nr}} \right)_{\rm flux} &=& N_T \, n_\chi \, \int  d^3 {\bm v} \,  f({\bm v}+{\bm v_E}) \nonumber\\
&\times& \left(\dfrac{d(\sigma v)}{dE_{\rm nr}} \right)_{\rm single} \, ,
\label{eq:differential scattering rate for a flux of dark matter particles}
\end{eqnarray}                                                                                                                                                                     
where ${\bm v}$ is the dark matter velocity in the Galactic frame, ${\bm v_E}$ is the average velocity of the Earth, $N_T$ is the number of target nucleus and $n_\chi$ is the number density of the constituents of dark matter.  We use the truncated Maxwell-Boltzmann distribution\,\cite{Lewin:1995rx} for the dark matter velocity distribution:
\begin{eqnarray}
f({\bm v} + {\bm v_E}) = N \, e^{- ({\bm v} + {\bm v_E})^2/2 v_0^2} \, \Theta(v_{\rm max} - v),
\label{eq:f(v+vE)}
\end{eqnarray}
with $v_E = 242 \,{\rm km \, s}^{-1}$, maximum velocity $v_{\rm max}$ = \mbox{600 km s$^{-1}$}, and velocity dispersion $v_0$ = \mbox{230 km s$^{-1}$}.  The normalization constant $N$ is adjusted so that $\int d^3 {\bm v} \, f({\bm v}+{\bm v_E})$ = 1.  Although recent simulations show that the velocity distribution of dark matter particles can be different from what we have assumed\,\cite{Mao:2012hf}, we use the truncated Maxwell-Boltzmann distribution as a proof of concept because of its simplicity.

The nuclear recoil energy spectrum that will be observed in a dark matter direct detection experiment is determined by how the full local dark matter density is distributed between dark matter particles and the darkonium.  We will contrast the nuclear recoil spectra for the two extreme situations:  
\begin{itemize}
\item all dark matter is composed of individual particles with mass $m$, 
\item all dark matter consists of darkonium bound states with approximate mass $2 m$.
\end{itemize} 
The local mass density of dark matter is $\rho_{\chi}$ = 0.3 GeV cm$^{-3}$.  If the  dark matter is fully made up of dark matter particles, the local number density of dark matter particles is $n_\chi = \rho_\chi/ m$.  If the dark matter is fully made up of darkonium, the local number density is $n_\chi = \rho_\chi/ 2 m$.

The universal two-body bound states that we are considering for the internal structure of dark matter are motivated by the large elastic cross sections proposed to solve small scale structure problems in $\Lambda$CDM.  We therefore determine the inverse scattering length $\gamma$ by taking the elastic self-interaction cross section per unit mass for dark matter particles to be $\sigma_{\rm el}/ m = 1$ cm$^2$ g$^{-1}$ at $v = 10$ km s$^{-1}$.  This corresponds to a binding energy $\gamma^2/m$ of 54 keV for $m = 10$ GeV and 0.52 keV for $m = 100$  GeV.  Much larger elastic cross sections, which corresponds to much smaller binding energies, are not allowed from cluster observations\,\cite{Clowe:2006eq}.  However, much smaller elastic cross sections which will have no effect on astrophysical scales are allowed, and those will correspond to much larger binding energy of the darkonium, so that the recoil rate of the darkonium breakup is further suppressed in that case.  Direct detection of self-interacting dark matter in a different particle physics model context is also presented in\,\cite{Kaplinghat:2013yxa}.  

The various recoil spectra for the differential event rate are shown in Fig.\,\ref{fig:Recoil spectra}.  Due to the very different masses of the dark matter candidates considered, and due to the variety of target nuclei considered, the scales in the $y$-axes of Fig.\,\ref{fig:Recoil spectra} vary.  In each plot we show the recoil energy spectra of the target nucleus for dark matter particle scattering,  darkonium elastic scattering, darkonium break up scattering, and total darkonium scattering.  For $m= 100$ GeV, at low recoil energies, the differential darkonium elastic scattering rate is approximately double the differential particle scattering rate.  This can be intuitively understood as the effect of the heavier incoming mass of the darkonium.  At low recoil energies, the form factor of the darkonium is almost one and hence the differential recoil rate of the darkonium elastic scattering is two times the differential recoil rate of a dark matter particle scattering.  A factor of four enhancement due to the coherent scattering of the darkonium is reduced by a factor of two due to the lower number of darkonium compared to the elementary dark matter particles for a given local dark matter density.   At higher recoil energies, the differential elastic scattering rate falls faster for darkonium than for a dark matter particle at higher energies due to the additional form factor suppression of the darkonium.  

We next compare the nuclear recoil energy spectrum from darkonium breakup.  For $m = 100$ GeV, the nuclear recoil energy spectrum vanishes at low nuclear recoils, peaks at a nuclear recoil energy that depends on the target nucleus and the binding energy of the darkonium, and subsequently falls much more slowly than that for darkonium elastic scattering case.  The vanishing of the nuclear recoil energy spectrum at zero recoil energies for the case of darkonium breakup is expected as a nonzero nuclear recoil is required to break up the darkonium.  Overall, the total nuclear recoil energy spectrum for an incident darkonium particle, which is the sum of the contribution of both the darkonium elastic scattering and darkonium breakup, is different from that of an incident dark matter particle both in shape and normalization.  

The total recoil energy spectrum from darkonium scattering looks similar to that of a dark matter particle of mass $2m$ with a $\sigma_{\rm SI}$ which is 4 times larger than for the other lines in the figure ($i.e.$, $\sigma_{\rm SI} = 10^{-42}$ cm$^2$).  If the dark matter mass is not known, then this degeneracy will be difficult to differentiate with low statistics.  If the dark matter mass is known via other measurements, then the end point in the nuclear recoil energy spectrum will determine whether the incident dark matter is a darkonium or a dark matter particle.  However, with high statistics, the differences in the nuclear recoil energy spectrum between that of an incident darkonium and an incident dark matter particle with an enhanced coupling to nucleons can be distinguished.  
  
For the $m = 10$ GeV case, due to the lower mass of the incident darkonium, the darkonium breakup is either extremely suppressed or kinematically forbidden.  It is therefore not visible in Fig.\,\ref{fig:Recoil spectra}.  Similar to the previous case, at low recoil energies the differential elastic nuclear recoil rate is approximately twice for an incident darkonium compared to that of an incident dark matter particle.  At larger recoil energies, the nuclear recoil energy spectrum for an incident darkonium decreases more slowly than that for an incident elementary dark matter particle.  The effect of the form factor is relatively small.  At the highest nuclear recoil energies shown, the form factor decreases the rate only by $\sim$ 20\%.  Even for light dark matter, the recoil energy spectrum looks similar to that of a dark matter particle of mass $2m$ with a $\sigma_{\rm SI}$ which is 4 times larger than for the other lines in the figure ($i.e.$, $\sigma_{\rm SI} = 10^{-37}$ cm$^2$).  This degeneracy can be broken either with information from other experiments or with high statistics.

For both masses, the total recoil energy spectrum from darkonium scattering from a nucleus is completely different from that for a single dark matter particle.  It is closer to the recoil energy spectrum for a dark matter particle with twice the mass and 4 times the cross section with a nucleon, but the shape is different.  The difference in shape is due to the form factor of the darkonium and to the new scattering channel in which the darkonium breaks apart.  We do not know of any another physical phenomenon which can give rise to such a different nuclear recoil energy spectrum.  

%%%%%%%%%%%%%%%%%%%%%%%%%%%%%%%%%%%%%%%%%%%%%%%%%%%
\section{Conclusion}
\label{sec:conclusion}
%%%%%%%%%%%%%%%%%%%%%%%%%%%%%%%%%%%%%%%%%%%%%%%%%%%

We have discussed the prospects of direct detection of dark matter with internal structure in the context of self-interacting asymmetric dark matter.  Our basic assumption is motivated by the possibility that large self-interaction cross sections for dark matter at nonrelativistic velocities can solve small-scale structure problems.  The assumption is that there is an energy region in which the cross section for a pair of dark matter particles come close to saturating the S-wave unitarity bound.  In this case, dark matter at lower energies has universal behavior that is completely determined by the S-wave scattering length.  The assumption requires that a pair of dark mater particles have an S-wave resonance near the scattering threshold.  If the resonance is just below the scattering threshold, it is a bound state of the two dark matter particles (we call it darkonium).  If the dark matter is asymmetric, darkonium can be stable and make up some or all of the present dark matter.  Due to the large scattering length, both the self-interaction cross section and the binding energy of the darkonium are determined by a single real parameter.  

Our assumption is predictive, because it implies that darkonium has universal low-energy properties that are completely determined by the scattering length.  In particular, the scattering length determines the shape of the cross sections for scattering of darkonium from a nucleus at sufficiently low recoil energy.  This implies new signatures that can be seen in a dark matter direct detection experiment, particularly for $\sim$ 100 GeV dark matter.  If a darkonium is incident on a target nuclei, two different final states are possible: (a) elastic scattering and (b) inelastic scattering where the darkonium breaks up from scattering with target nuclei.   Due to the extended spatial structure of the darkonium and the possibility of breakup, the nuclear recoil energy spectrum in a dark matter direct detection experiment will be different from that due to an incident dark matter particle.  Some examples of the nuclear recoil energy spectrum due to an incident darkonium are shown in Fig.\,\ref{fig:Recoil spectra}.  As can be seen from the figure, the total nuclear recoil energy spectrum due to an incident darkonium is completely different from that due to an incident dark matter particle.  It is similar to the recoil energy spectrum for a dark matter particle with twice the mass and 4 times the cross section with a nucleon, but there is a difference in the shape.  If a nuclear recoil spectrum of this kind is unambiguously seen in dark matter direct detection experiments, then it will be a smoking-gun signature for internal structure in dark matter.  

\section*{Acknowledgments} 

We especially thank John Beacom for discussions which led to this idea and for extensive discussions during this work.  We thank Kenny C. Y. Ng and Annika Peter for discussions and clarifications.  R.\,L. is supported by NSF Grant PHY-1101216 to John Beacom.  E.\,B. is supported by the Department of Energy under grant DE-FG02-ER40690.

%%%%%%%%%%%%%%%%%%%%%%%%%%%%%%%%%%%%%%%%%%%%%%%%%%%
\section*{Appendix}
\label{sec:appendix}
%%%%%%%%%%%%%%%%%%%%%%%%%%%%%%%%%%%%%%%%%%%%%%%%%%%

In this appendix, we present the detailed derivation of the recoil energy spectrum of a nucleus in a dark matter direct detection experiment.  We begin by presenting the Feynman rules that are used for the derivation.  We derive the nuclear recoil energy spectrum first for a dark matter particle scattering off a nucleus and then for a bound state of two dark matter particles (darkonium) scattering off a nucleus.  For a darkonium scattering off a nucleus, there are two possible final states: (a) the darkonium is still bound after the scattering, and (b) the darkonium is broken apart due to the scattering. 

\subsection*{Feynman Rules}

Particles with a large scattering length can be described by a renormalizable local quantum field theory.  The Feynman rules for the quantum field theory are simple\,\cite{Braaten:2004rn}.  The particles have standard nonrelativistic propagators.  A pair of particles can interact through a point interaction vertex with a bare coupling constant $g_0$.  They can rescatter through additional interaction vertices.  The resulting bubble diagrams are ultraviolet divergent and require an ultraviolet cutoff $\Lambda$.  The interaction is nonperturbative, so the bubble diagrams must be summed up to all orders.  The scattering amplitude for a pair of particles is the sum of arbitrarily many bubble diagrams.  Renormalization is implemented by tuning the bare coupling constant as a function of $\Lambda$ so that the inverse scattering length has the desired value $\gamma$.  Amplitudes in this quantum field theory can be calculated more easily by using a more succinct set of Feynman rules in which arbitrarily many bubble diagrams have been summed up to all orders.  Renormalisation allows these Feynman rules to be expressed in terms of the physical parameter $\gamma$.

The Feynman rules for identical bosons, which are illustrated in Fig.\,\ref{fig:Feynman rules}\,\cite{Braaten:2004rn}, involve the following factors:

\begin{itemize}
\item The nonrelativistic propagator for a virtual dark matter particle of energy $E$ and momentum ${\bm p}$ is given by  $i D(E,{\bm p})$, where
\begin{eqnarray}
D(E,{\bm p}) = \dfrac{1}{E - {\bm p}^2/2m + i \epsilon}\,.
\label{eq:propagator}
\end{eqnarray}
It is represented by a single dashed line.  

\item The product of the residue factor for an incoming darkonium line and the vertex factor for its transition to a pair of particles is given by $- i g_2$, where
\begin{eqnarray}
g_2 = \sqrt{\dfrac{16 \pi \gamma}{m^2}}\,.
\label{eq:g2}
\end{eqnarray}
It is represented by a dot at which a double-dashed line splits into two dashed lines as shown in Fig.\,\ref{fig:Feynman rules}.  Since bubble diagrams have already been summed up to all orders, the first interaction of the pair of particles cannot be with each other.

\begin{figure}[!h]
\centering
\includegraphics[angle=0.0,width=0.48\textwidth]{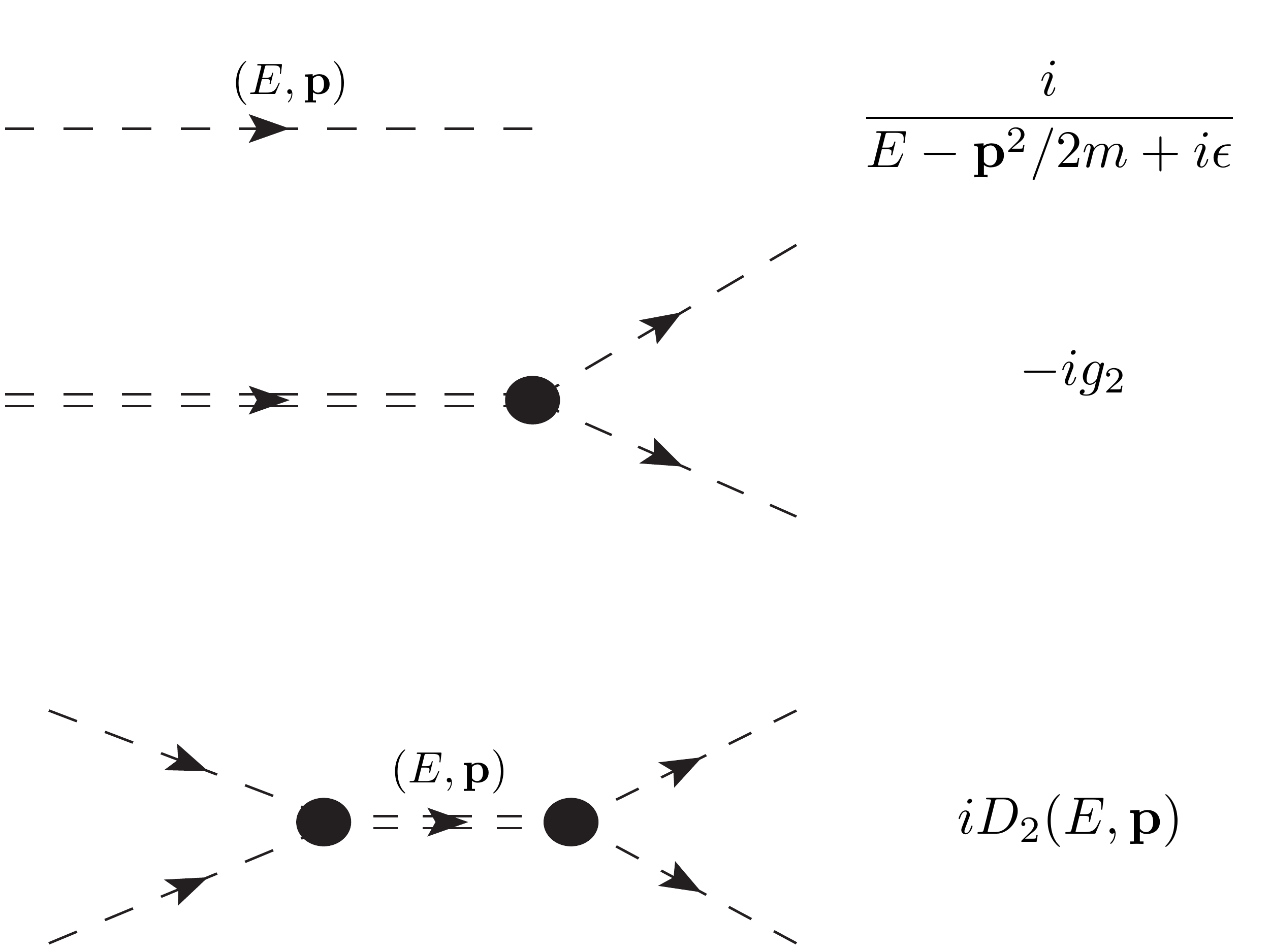}
\caption{Feynman rules for the particle propagator, the product of the residue and vertex factors for an incoming darkonium, and the 2$\rightarrow$2 transition amplitude for a pair of particles.}
\label{fig:Feynman rules}
\end{figure}

\item The exact 2$\rightarrow$2 transition amplitude for a pair of particles with total energy $E$ and total momentum $\bm p$ is given by $i D_2(E,{\bm p})$, where
\begin{eqnarray}
D_2(E,{\bm p}) = \dfrac{8 \pi/m}{-\gamma + \sqrt{-m(E - {\bm p}^2/4m + i \epsilon)}} \, . \phantom{11111}
\label{eq:D2}
\end{eqnarray}
It is represented by a double dashed line joined by two dots as shown in Fig.\,\ref{fig:Feynman rules}.  The first previous interaction of the incoming pair of particles cannot be with each other.  The first subsequent interaction of the outgoing pair of particles cannot be with each other.

\item The vertex factor for the scattering of a dark matter particle from the nucleus with momentum transfer $q$, as in the diagram in Fig.\,\ref{fig:Feynman diagram particle scattering}, is given by $- i \,
G_A(q)$.

\end{itemize}

If $\gamma >0$, the amplitude $D_2(E,{\bf p})$ in Eqn.\,(\ref{eq:D2}) has a pole in the energy at ${\bf p}^2/4m - \gamma^2/m$.  This corresponds to a darkonium with momentum ${\bf p}$ and binding energy $\gamma^2/m$, in accord with Eqn.\,(\ref{eq:real gamma binding energy}).  Up to a complex phase, the product $g_2$ of the residue factor and the vertex factor in Eqn.\,(\ref{eq:g2}) is the square root of the residue of that pole.  The matrix element for scattering of a pair of particles with momenta $+{\bf k}$ and $-{\bf k}$, which implies total energy $k^2/m$ is 
\begin{eqnarray}
D_2(k^2/m,0) = \dfrac{8 \pi}{-\gamma - ik} \, .
\label{eq:D2k^2/m,0}
\end{eqnarray}
The cross section is $|D_2(k^2/m,0)|^2$ multiplied by the flux factor $m/2k$ for the incoming particles and by the phase space factor $mk/4 \pi$ for the outgoing identical particles.  This reproduces the elastic cross section in Eqn.\,(\ref{eq:real gamma sigma elastic}).

The Feynman rules given above are for the case in which the particles with the large scattering length are identical bosons.  If the particles are distinguishable, they can have distinct masses $m_1$ and $m_2$.  Their propagators are obtained by replacing $m$ in Eqn.\,(\ref{eq:propagator}) by $m_1$ or $m_2$.  The exact 2$\rightarrow$2 transition amplitude is obtained from Eqn.\,(\ref{eq:D2}) by replacing 8$\pi$ by 4$\pi$ and by replacing the mass $m$ by 2$\mu_{12}$, where $\mu_{12}$ is the reduced mass of the two particles.  The product of the residue factor and the vertex factor is obtained from Eqn.\,(\ref{eq:g2}) by making the same two replacements, resulting in $(\pi \gamma/ \mu_{12}^2)^{1/2}$.  The two particles can scatter from a nucleus with different amplitudes $G_{A,1}(q)$ and $G_{A,2}(q)$.

The Feynman diagrams for scattering of a dark matter particle and of a darkonium from a nucleus are shown in Fig.\,\ref{fig:Feynman diagram particle scattering} and in Figs.\,\ref{fig:Feynman diagram darkonium scattering} and \ref{fig:Feynman diagram darkonium breakup} respectively.  We denote the incoming momentum of the dark matter particle or the darkonium by $\bm{P}$ and the momentum of the target nucleus by $\bm{K}$.  The total momentum of the outgoing dark matter, which can be a single particle or a darkonium is denoted by ${\bm P'}$.  For darkonium breakup, the momenta of the two outgoing dark matter particles are denoted by ${\bm p}_1$ and ${\bm p}_2$ respectively.  The momentum of the scattered nucleus is denoted by $\bm{K'}$.  In the laboratory frame, the target nucleus is almost at rest, so $\bm{K}$ = 0 to a very good approximation.  The momentum transferred to the nucleus by the scattering is $\bm{q} = \bm{K'} - \bm{K}$.  The momentum transfer ${\bm q}$ is independent of the Galilean frame and its magnitude is denoted by $q$.  The recoil energy of the scattered nucleus in the laboratory frame is $E_{\rm nr} = q^2/2 m_A$.  

\subsection*{Scattering of dark matter particle}

In this section, we will detail the recoil energy spectrum of the scattered nucleus due to scattering with a dark matter particle.  The nonrelativistic phase space in a general Galilean frame is denoted by 
\begin{eqnarray}
\left(d\Phi \right)_{A+1} &=& \dfrac{d^3 {\bm {P'}}}{(2 \pi)^3} \dfrac{d^3 {\bm {K'}}}{(2 \pi)^3} (2 \pi)^3 \, \delta ^3 ({\bm {P + K - P' - K'}}) \nonumber\\
&&\times  \, 2\pi \, \delta \left(\dfrac{{\bm P}^2 - {\bm P'}^2}{2 m} + \dfrac{{\bm K}^2-{\bm K'}^2}{2 m_A}\right) \,. \phantom{111111}
\label{eq:non-relativistic 2-body phase space}
\end{eqnarray}
In the laboratory frame, ${\bm K} = 0$ and the momentum transfer reduces to ${\bm q} = {\bm K'}$.  The phase space can be simplified to 
\begin{eqnarray}
\left( d\Phi \right)_{A+1,{\rm lab}} = \dfrac{q^2 dq \, d ({\rm cos}\theta)}{2 \pi} \dfrac{m}{P q} \, \delta \left({\rm cos}\theta - \dfrac{m q}{2 \, \mu \, P}\right) \, ,\phantom{1111}
\label{eq:non-relativistic 2-body phase space simplified}
\end{eqnarray}
where $\theta$ is the angle between $\bm{q}$ and $\bm{P}$ and $\mu$ is the reduced mass of the dark matter particle and the nucleus.  The delta function determines the minimum velocity $v = P/m$ of the dark matter particle to produce a recoil of momentum $q$: $v \geq q/ 2 \mu$.

The Feynman diagram for the scattering of a dark matter particle from the nucleus is shown in Fig.\,\ref{fig:Feynman diagram particle scattering}.  The matrix element for the process is $- \, G_{A}(q)$.  The differential scattering rate $v d\sigma$ is $|G_A(q)|^2$ multiplied by the differential phase space in Eqn.\,(\ref{eq:non-relativistic 2-body phase space simplified}).  After integrating over the scattering angle, we obtain Eqn.\,(\ref{eq:differential scattering rate wimp}).

\subsection*{Elastic scattering of darkonium}

In this section, we detail the recoil energy spectrum of a nucleus due to elastic scattering of a bound state of dark matter (darkonium) off the target nucleus.  The nonrelativistic phase space is similar to Eqn.~(\ref{eq:non-relativistic 2-body phase space}), except that the mass $m$ of the dark matter is replaced by $2 m$.  In the laboratory frame, the phase space can be simplified to give
\begin{eqnarray}
\left(d \Phi \right)_{A+2} = \dfrac{q^2 dq \,d ({\rm cos} \theta)}{2 \pi} \dfrac{2 m}{P q} \delta \left({\rm cos}\theta - \dfrac{m q}{\mu_2 P}\right)  \, ,\phantom{1111}
\label{eq:non-relativistic 2-body phase space simplified darkonium}
\end{eqnarray}
where $\mu_2$ is the reduced mass of the darkonium and the nucleus.  The delta function determines the minimum velocity $v = P/ 2m$ of darkonium necessary to produce a nuclear recoil of momentum $q$: $v \geq q/(2 \mu_2)$.

The Feynman diagram for the process is shown in Fig.\,\ref{fig:Feynman diagram darkonium scattering}.  The matrix element is given by 
\begin{eqnarray}
\mathcal{M} &=& - i \, G_{A}(q) \, g_2^2 \int \dfrac{d^3 k}{(2 \pi)^3} \int \dfrac{d \omega}{2 \pi} \, D(\omega, {\bm k}) \nonumber\\
&\times& D (- E_B + P^2/4m - \omega, {\bm P} - {\bm k})\nonumber\\
&\times& D(-E_B+ P'^2/4m - \omega, {\bm P}' - {\bm k})  \,, 
\label{eq:matrix element darkonium}
\end{eqnarray}
where ${\bm k}$ and $\omega$ are the undetermined momentum and energy in the loop.  The integral over $\omega$ can be evaluated by closing the contour in the lower half-plane around the pole of $D(\omega, {\bm k})$.  The integral over ${\bm k}$ can be evaluated after combining the remaining two propagator using a Feynman parameter.  Upon integrating over the Feynman parameter, the matrix element reduces to
\begin{eqnarray}
&& \mathcal{M} = - G_A(q) g_2^2 \, \dfrac{m^2}{2 \pi q} \, {\rm tan}^{-1} \dfrac{q}{4 \gamma}\, . \phantom{1111}
\label{eq:darkonium scattering simplified}
\end{eqnarray}
The differential rate $v\, d\sigma$ for elastic scattering of a darkonium of momentum $P \approx 2 m \, v$ is obtained by squaring the matrix element and multiplying by the differential phase space in Eqn.\,(\ref{eq:non-relativistic 2-body phase space simplified darkonium}).  After integrating over the scattering angle, we obtain the differential scattering rate in Eqn.\,(\ref{eq:differential rate of darkonium scattering}).

We now consider the case in which the constituents of darkonium have the same mass $m$ but different amplitudes for scattering from the nucleus.  In this case, there are two diagrams like the one in Fig.\,\ref{fig:Feynman diagram darkonium scattering} with different vertex factors $G_{A,1}(q)$ and $G_{A,2}(q)$.  Because the particles are distinguishable, the factor $g_2$ for an external darkonium line is smaller than that in Eqn.\,(\ref{eq:g2}) by a factor of 2.  The net effect on the final expression for the differential scattering rate in Eqn.\,(\ref{eq:differential rate of darkonium scattering}) is that $G_A(q)$ is replaced by $[G_{A,1}(q) +G_{A,2}(q)]/4$.  It reduces to Eqn.\,(\ref{eq:differential rate of darkonium scattering}) if we set $G_{A,1}(q) = G_{A,2}(q) = 2 \, G_A(q)$.

\subsection*{Breakup scattering of darkonium}

Here we detail the recoil energy spectrum of a bound state of two dark matter particles (darkonium) breaking apart after scattering from the nucleus.  We denote the momenta of the two outgoing dark matter particles by ${\bm p_1}$ and ${\bm p_2}$.  The nonrelativistic phase space in a general Galilean frame is given by 
\begin{eqnarray}
&&\left(d\Phi \right)_{A+1+1} = \dfrac{d^3 {\bm p}_1}{(2 \pi)^3} \dfrac{d^3 {\bm p}_2}{(2 \pi)^3} \dfrac{d^3 {\bm {K'}}}{(2 \pi)^3} \nonumber\\
&\times& (2 \pi)^3 \, \delta ^3 ({\bm P} + {\bm K} - {\bm p}_1 - {\bm p}_2 - {\bm K'}) \nonumber\\
&\times& 2\pi \, \delta \left( \dfrac{{\bm P}^2- 2 ({\bm p}_1^2 + {\bm p}_2^2)}{4 m} - E_B + \dfrac{{\bm K}^2 - {\bm K'}^2}{2 m_A}\right). \phantom{1111}
\label{eq:darkonium breakup phase space}
\end{eqnarray}
We employ  the change of variables ${\bm p}_{1,2} = \frac 12 \,{\bm P'} \pm {\bm r}$ and use the delta function to integrate over ${\bm P'}$.  In the laboratory frame, the phase space can be reduced to 
\begin{eqnarray}
\left(d\Phi \right)_{A+1+1,{\rm lab}} &=& \dfrac{d^3 {\bm q}}{(2 \pi)^3} \,\dfrac{d^3 {\bm r}}{(2 \pi)^3}   \nonumber\\
&\times& 2 \pi \, \delta \left( \dfrac{{\bm P}\cdot{\bm q} - 2 {\bm r}^2}{2 m} - E_B - \dfrac{\bm q^2}{2 \mu_2} \right) \, .\phantom{11}
\label{eq:darkonium breakup phase space simplified}
\end{eqnarray}

The Feynman diagrams for the breakup of darkonium from the scattering of the nucleus are shown in Fig.\,\ref{fig:Feynman diagram darkonium breakup}.  The matrix element is the sum of three terms.  The matrix elements for the first diagram in Fig.\,\ref{fig:Feynman diagram darkonium breakup} and for the diagram obtained by interchanging ${\bm p_1}$ and ${\bm p_2}$ are
\begin{eqnarray}
\mathcal{M}_{1} &=& \dfrac{4 m \, g_2 \, G_A(q)}{4 \gamma^2 + (2 {\bm r} - {\bm q})^2}\, , \label{eq:scattering amplitude darkonium breakup M1_1}\\
\mathcal{M}_2 &=& \dfrac{4 m \, g_2 \, G_A(q)}{4 \gamma^2 + (2 {\bm r} + {\bm q})^2}\, .
\label{eq:scattering amplitude darkonium breakup M1}
\end{eqnarray}
The matrix element for the second Feynman diagram in Fig.\,\ref{fig:Feynman diagram darkonium breakup} can be written as
\begin{eqnarray}
\mathcal{M}_3 &=& i \, G_{A}(q) \, g_2 \dfrac{8 \pi/m}{-\gamma - i r}\,\int \dfrac{d^3 k}{(2 \pi)^3} \int \dfrac{d \omega}{2 \pi} D(\omega,{\bm k}) \nonumber\\ 
&\times&D(-E_B + {\bm P}^2/ 4 m - \omega, {\bm P} - {\bm k}) \nonumber\\
&\times&D(({\bm p}_1^2 + {\bm p}_2^2)/ 2 m - \omega, {\bm p}_1+{\bm p}_2 - {\bm k}) \, . 
\label{eq:scattering amplitude darkonium breakup}
\end{eqnarray}
The integral over $\omega$ can be evaluated by closing the contour in the lower half-plane, so that it encloses the pole of $\omega$.  The integral over ${\bm k}$ can be evaluated after combining the remaining two propagators with a Feynman parameter.  The matrix element in Eqn.\,(\ref{eq:scattering amplitude darkonium breakup}) reduces to 
\begin{eqnarray}
\mathcal{M}_3 = - \dfrac{2 i \, m \,  g_2 \, G_A(q) \,}{q \, (\gamma + i r)} \, {\rm ln} \dfrac{4 r^2 + (2 \gamma - i q)^2}{4 \gamma^2 + (q - 2 r)^2} \, . \phantom{111}
\label{eq:darkonium breakup simplified}
\end{eqnarray}

The complete matrix element is $\mathcal{M}_1+\mathcal{M}_2+\mathcal{M}_3$.  The differential rate $v \, d\sigma$ for the breakup scattering of a darkonium of momentum $P \approx 2 m \, v$ is obtained by squaring the matrix element and multiplying by the differential phase space in Eqn.\,(\ref{eq:darkonium breakup phase space simplified}).  Now $|\mathcal{M}|^2$ depends only on the angle between ${\bm r}$ and ${\bm q}$, and the argument of the delta function depends only on the angle between ${\bm q}$ and ${\bm P}$.  After averaging $|\mathcal{M}|^2$ over the angles of ${\bm r}$, we can use the delta function to evaluate the angular integral for ${\bm q}$.  The compact expression for the differential rate in Eqn.\,(\ref{eq:darkonium breakup phase space simplified compact}) is obtained by subsequently reexpressing the angle average of $|\mathcal{M}|^2$ in terms of an integral over the angles of ${\bm r}$.

We now consider the case in which the constituents of darkonium have the same mass $m$ but different amplitudes for scattering from the nucleus.  Because the particles are distinguishable, the factor $g_2$ for an external darkonium line is smaller than that in Eq. (17) by a factor of 2.  The $2 \to $2 transition amplitude is also smaller than that in Eqn.\,(\ref{eq:D2}) by a factor of 2.  The effect on the matrix element is to replace $G_A(q)$ in Eqns.\,(\ref{eq:scattering amplitude darkonium breakup M1_1}), (\ref{eq:scattering amplitude darkonium breakup M1}) and (\ref{eq:scattering amplitude darkonium breakup}) by $G_{A,1}(q)$, $G_{A,2}(q)$, and $[G_{A,1}(q) + G_{A,2}(q)]/2$,
respectively.  The final expression for the differential scattering rate reduces to
Eqn.\,(\ref{eq:darkonium breakup phase space simplified compact}) if we set $G_{A,1}(q) = G_{A,2}(q) = 2 \, G_A(q)$.

\bibliographystyle{kp}
\interlinepenalty=10000
\tolerance=100
\bibliography{Bibliography/references}

\end{document}